\documentclass[twocolumn,superscriptaddress,showpacs]{revtex4-1}

\usepackage{amssymb}
\usepackage{amsmath}
\usepackage{graphicx}

\usepackage{epsf}
\usepackage{tikz}
\usepackage{epsfig}
\usepackage{epstopdf}
\usepackage{color}
\usepackage{color}
\usepackage{float}
\usepackage{comment}

\def\g{{\bf{g}}}

\def\p{{\bf{p}}}

\def\g0{{\gamma_0}}

\def\figwidth{0.9\columnwidth}

\begin{document}
\title{Dynamics of a Mobile Impurity in a Two Leg Bosonic Ladder}
\author{Naushad Ahmad Kamar}
\affiliation{DQMP, University of Geneva, 24 Quai Ernest-Ansermet, 1211 Geneva, Switzerland}
\author{Adrian Kantian}
\affiliation{Department of Physics and Astronomy, Uppsala University, Box 516, S-751 20 Uppsala, Sweden}
\author{Thierry Giamarchi}
\affiliation{DQMP, University of Geneva, 24 Quai Ernest-Ansermet, 1211 Geneva, Switzerland}
\begin{abstract}
We have analyzed the behavior of a mobile quantum impurity in a bath formed by a two-leg bosonic ladder by a combination of field theory (Tomonaga-Luttinger liquid) and numerical (Density Matrix Renormalization Group) techniques. 
Computing the Green's function of the impurity as a function of time at different momenta, we find a power law decay at zero momentum, which signals the breakdown of any quasi-particle description of the impurity motion. 
We compute the exponent both for the limits of weak and strong impurity-bath interactions. 
At small impurity-bath interaction, we find that the impurity experiences the ladder as a single channel one-dimensional bath, but effective  coupling is reduced by a factor of $\sqrt 2$, thus the impurity is less mobile in the ladder 
compared to a one dimensional bath. We compared the numerical results for the exponent at zero momentum with a semi-analytical expression that was initially established for the chain and find excellent agreement without adjustable parameters. 
We analyze the dependence of the exponent in the transverse hopping in the bath and find surprisingly an increase of the exponent at variance with the naive extrapolation of the single channel regime. 
We study the momentum dependence of the impurity Green's function and find that, as for the single chain, two different regime of motion exist, one dominated by infrared metatrophy and a more conventional polaronic behavior.
We compute the critical momentum between these two regimes and compare with prediction based on the structure factor of the bath. In the polaronic regime we also compute numerically the lifetime of the polaron. 
Finally we discuss how our results could be measured in cold atomic experiments. 
\end{abstract}
\maketitle

\section{Introduction}

How a quantum environment can affect the properties of a quantum particle is one of the central problems for quantum many-body systems. In a solid, where the bath is formed by  phonon excitations, this is at the heart of the polaron problem~\cite{feynman55_polaron,feynman_statmech}.
For interacting particles, the dressing of one particle by the excitations of the density induced by the interactions leads to the
Fermi liquid description of an interacting fermion gas~\cite{landau_fermiliquid_theory_static,nozieres_book}. In addition to these types of effects, it was realized by Caldeira and Leggett that the environment can deeply change the properties
of a quantum degree of freedom in a way that goes beyond the formation of a polaron with its attendant redefinition of some physical parameters of the particle, such as the mass~\cite{caldeira_leggett,leggett_two_state}.

One potentially important ingredient is the dimensionality of system. It has been shown~\cite{zvonarev_ferro_cold} that for a one-dimensional bath the influence on the propagation of the particle
can be drastic. The particle motion may become subdiffusive with no remnant of quasiparticle behavior. The source for the onset of this new dynamical universality class in the one-dimensional systems is a phenomenon akin to the Anderson orthogonality catastrophe. Following this, the phenomenon received more extensive study~\cite{kamenev_exponents_impurity,matveev_exponent_ferro_liquid,schecter_mobile_impurity,Lamacraft_mobile_impurity_in_one_dimension,zvonarev_bosehubbard,zvonarev_gaudinyang}. Combining numerical and analytical results~\cite{kantian_impurity_DMRG} even showed that depending on the momentum
of the particle a change in regime, from subdiffusive to polaronic, could occur. Extensions of these results to fermionic systems~\cite{massel_kicked_impurity,horovitz_impurity_fermions}, driven particles \cite{mathy_supersonic_impurity,knap_impurity_flutter} or particles coupled to several one dimensional systems \cite{horovitz_impurity_many_TLL_transconductance} have since been performed.

On the experimental front, cold atomic systems have brought an unprecedented level of control to study these issues~\cite{Michael_Kohl_Spin_impurity_in_bose_gas,Minardi_Dynamics_of_impurities_in_one_dimension,Fukuhara_spin_impurity_in_one_dimension,meinert_bloch_oscillations_TLL}, since they allow for excellent control of the interaction
between the impurity and the bath, as well as direct measurements of correlation functions.

How the breakdown of the quasiparticle picture and the emergence of subdiffusive dynamics would be affected if the dimensionality of the bath increases is an open and interesting question. Indeed, in other effects
in which orthogonality manifests, such as the X-ray edge problem~\cite{Giamarchi_Bosonization}, it is well known that recoil of the impurity suppresses the orthogonality when the dimension is larger than one for the bath. It is thus natural to wonder
for the problem of the mobile impurity how one actually crosses over from the singular physics of a one dimensional bath to the more conventional polaronic dynamics~\cite{grusdt_polaron_review_cold} that one could naively expect for two or three dimensional baths.

In this paper we address this issue by looking at an impurity propagating in a ladder. We still confine the impurity to move one-dimensionally, but the bath has now a transverse extension. We mostly focus on the case of a two leg ladder but will also
discuss some of the consequences of increasing the number of legs.
We analyse this problem using a combination of analytical and numerical techniques in a spirit similar to the one used for the single chain~\cite{kantian_impurity_DMRG}, and compare the results of the ladder bath to the known results for the single chain.

The plan of the paper is as follows: Sec.~\ref{sec:definitions} presents the model, the various observable studied and the bosonization
representation that will be at the heart of the analytical analysis. Sec.~\ref{sec:analytic} presents the two main analytical approaches
that are used, namely a field theory representation based on the bosonization technique and a linked cluster expansion representation.
Sec.~\ref{sec:numerics} presents the numerical Density Matrix Renormalization Group (DMRG)\cite{white_dmrg_letter,Schollwock_DMRG} analysis of this problem, and the results for the Green's function of the impurity. Sec.~\ref{sec:discussion} discusses these results both in connection with the single chain
results and in view of the possible extensions. Finally Sec.~\ref{sec:conclusion} concludes the paper and presents some perspectives in connection with experiments. Technical details can be found in the Appendices.

\section{Mobile Impurity in a Two Leg Bosonic Ladder} \label{sec:definitions}

\subsection{Model} \label{sec:model}

We consider a mobile impurity moving in a two-leg Bosonic ladder. In this study we restrict the motion of the impurity to be strictly one-dimensional, other cases will be considered elsewhere~\cite{kamar_impurity_twolegs}. This model is thus the simplest basic model for exploring the dynamics of a one-dimensional impurity in an environment that is no longer purely 1D itself.

\begin{figure}
\begin{center}
 \includegraphics[width=\figwidth]{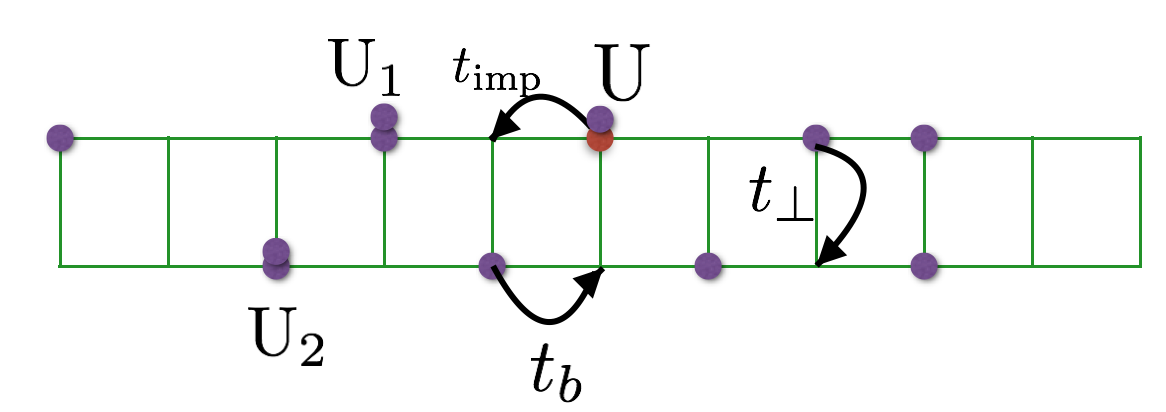}
\caption{\label{fig:2fig1}(color online) Impurity in a two leg  bosonic ladder. Magenta solid circles represent the bath particles and the red circle represents the impurity.
 The bath particles move along the legs (between the legs) with hopping $t_b$ ($t_\perp$) and have a contact interaction $U_1$ ($U_2$) between themselves (see text).
 The impurity motion is restricted to the upper leg, its amplitude is $t_{\text{imp}}$ (see text). The impurity and the bath particles interact by a contact interaction $U$.}
\end{center}
\end{figure}

The model we consider is depicted in Fig.~\ref{fig:2fig1}.
The full Hamiltonian is given by
\begin{equation} \label{eq:hamdiscrete}
 H  = H_K + H_{\text{lad}} + U \sum_j \rho_{1,j} \rho_{\text{imp},j}
\end{equation}
$U$ is the interaction strength between the particles in the ladder and the impurity.

The impurity kinetic energy is given by the tight-binding Hamiltonian
\begin{equation}
 H_K = - t_{\text{imp}} \sum_j (d^\dagger_{j+1} d_j + \text{h.c.})
\end{equation}
where $d_j$ ($d^\dagger_j$) are the destruction (creation) operators of the impurity on site $j$.
The density of the impurity on site $j$ is
\begin{equation}
 \rho_{\text{imp},j} = d^\dagger_j d_j
\end{equation}
The ladder Hamiltonian $H_{\text{lad}}$ is given by
\begin{equation} \label{eq:laddiscrete}
 H_{\text{lad}} = H^0_1 + H^0_2 - t_\perp \sum_j (b^\dagger_{1,j} b_{2,j} + \text{h.c.} )
\end{equation}
where $b_{a,j}$ ($b^\dagger_{a,j}$) are the destruction (creation) operators of a boson of the bath on chain $a$ and site $j$.
The $b$-operators obey the standard bosonic commutation relation rules.
The single chain Hamiltonian is the
Bose-Hubbard one
\begin{equation}
\begin{split}
 H^0_i &= -  t_b \sum_j (b^\dagger_{j+1} b_j + \text{h.c.}) + \frac{U_i}2 \sum_j \rho_{i,j}(\rho_{i,j} - 1) \\
& - \mu_i \sum_j \rho_{i,j}
 \end{split}
\end{equation}
In the following, we use the tunneling rate $t_b$ of the bath bosons as the unit of energy.

The form (\ref{eq:hamdiscrete}) is convenient for the numerical study. In order to make easily connection with the field theory analysis we can also consider the same problem
in a continuum. In this case the Hamiltonian becomes
\begin{equation} \label{eq:hamtot}
 H  = \frac{P^2}{2M} + H_{\text{lad}} + U \int d x \rho_1(x) \rho_{\text{imp}}(x)
\end{equation}
where $P$ and $M$ are respectively the momentum and mass of the impurity. The density of the impurity is
\begin{equation}\label{eq:imp_density}
 \rho_{\text{imp}}(x) = \delta(x - X)
\end{equation}
where $X$ is the position of the impurity, canonically conjugate to $P$ so that $[X,P] = i\hbar$.
We set from now on $\hbar=1$.

In the continuum, the ladder Hamiltonian (\ref{eq:laddiscrete}) becomes
\begin{equation} \label{eq:hamlad}
 H_{\text{lad}} = H^0_1 + H^0_2 - t_\perp \int dx (\psi^\dagger_1(x)\psi_2(x) + \text{h.c.})
\end{equation}
and the single chain Hamiltonian is
\begin{equation} \label{eq:hamchain}
\begin{split}
 H^0_i & = \frac1{2m} \int dx |\nabla\psi_i(x)|^2 + \frac{U_i}2 \int dx \rho_i(x)^2 \\
 & - \mu_i \int dx \rho_i(x)
 \end{split}
\end{equation}
$m$ is the mass of the bosons, $\mu_i$ is the chemical potential and $U_i$ is the intrachain interaction of leg $i=1,2$. $\psi(x)^\dagger$ ($\psi(x)$) is the creation (annihilation) operator at position $x$.

\subsection{Observables}

To characterize the dynamics of impurity in the ladder we mostly focus on the Green's function of the impurity. We study it both analytically and numerically via DMRG, each time considering the zero temperature case.

The Green's function of the impurity is defined as
\begin{equation} \label{eq:greenimp}
 G(p,t) = \langle \hat{d}_{p}(t) \hat{d}_p^\dagger(t=0) \rangle
\end{equation}
where $\langle \cdots \rangle$ denotes the average in the ground state of the bath, and with zero impurities present. $O(t)$ denotes the usual Heisenberg time evolution of the operator
\begin{equation}
 O(t) = e^{i H t} O e^{-i H t}
\end{equation}
and the operator $\hat{d}_p$ is the operator destroying an impurity with momentum $p$ given by
\begin{equation}
 \hat{d}_p = \sum_j e^{i p r_j} d_j
\end{equation}
with $r_j = a j$ on the lattice and the corresponding integral
\begin{equation}
 \hat{d}_p = \int dx e^{i p x} d(x)
\end{equation}
in the continuum.

\subsection{Bosonization representation}

To deal with the Hamiltonian defined in the previous section, it is convenient to focus on the relevant low-energy, long-wavelength properties using the so-called bosonization representation.
This representation is well-documented in the literature by now, and we recall here only the salient points to fix the notations.

The single-particle operators are represented in terms of two conjugate operators $\phi(x)$ and $\theta(x)$, capturing collective excitations of density and phase respectively, via the formulas~\cite{Giamarchi_Bosonization} .
\begin{equation}\label{eq:2eq1c}
 \rho_\alpha(x) =  \rho_{0,\alpha}-\frac{\nabla \phi_{\alpha}(x)}{\pi}+\rho_0 \sum_{p\neq 0} e^{2 i p(\pi\rho_{0, \alpha} x-\phi_{\alpha}(x))}
\end{equation}
where $\rho_{0,\alpha}$ is the average density on the chain $\alpha=1,2$. The creation operator of a particle in the bath in term of $\theta$ and $\phi$ is given to lowest order by
\begin{equation}\label{eq:2eq5}
 \psi_{\alpha}^{\dagger}(x) = \rho_{0,\alpha}^{1/2} e^{-i\theta_\alpha(x)}
\end{equation}
The conjugate field operators $\phi_{1,2}$ and $\theta_{1,2}$ obey
\begin{equation}\label{eq:2eq3}
[\phi(x_1),\frac{\nabla \theta(x_2)}{\pi}] = i \delta(x_1-x_2)
\end{equation}

Using the above representation, and assuming that the filling of the chains is not commensurate with the underlying lattice, allows rewriting the Hamiltonian of each chain as~\cite{Giamarchi_Bosonization}
\begin{equation}\label{eq:2eq2}
 H^0_\alpha = \frac{1}{2\pi}\int dx \left[u_\alpha K_\alpha(\partial_x \theta_\alpha)^2+\frac{u_\alpha}{K_\alpha} (\partial_x \phi_\alpha)^2 \right]
\end{equation}
where $u_\alpha$ and $K_\alpha$ are the so-called Tomonaga-Luttinger liquid (TLL) parameters. Here, $u_\alpha$ is the velocity of density excitations (i.e. sound) in the chain,
while $K_\alpha$ encodes the effect of interactions and controls the decay of the correlation functions. Their values can be directly related to the bare parameters of the given microscopic Hamiltonian \cite{Giamarchi_Bosonization,cazalilla_review_bosons}.
For example, for the Lieb-Lininger model \cite{lieb_bosons_1D} of bosons in the continuum with contact interaction
$K=\infty$ when the interaction is zero, and $K=1$ when the contact repulsion is infinite. For the two-leg bosonic ladder the TLL parameters can be found in \cite{crepin_bosonic_ladder_phase_diagram}.

Using the bosonization framework, the low-energy approximation of inter-leg tunneling is
\begin{equation}
 - 2 t_\perp \rho_0 \int dx\; \cos(\theta_1(x) - \theta_2(x)).
\end{equation}
This makes it convenient to use the symmetric and antisymmetric combinations of the fields
\begin{equation}
\begin{split}
 \theta_{s,a} &= \frac{\theta_1\pm\theta_2}{\sqrt{2}}
\end{split}
\end{equation}
and analogous expressions for the fields $\phi$. The new fields remain canonically conjugate and allow to re-express the Hamiltonian of the bath as
\begin{equation} \label{eq:ladcont}
 H_{\text{lad}} = H_s + H_a
\end{equation}
with
\begin{equation}\label{eq:2eq7}
\begin{split}
H_s =& \frac{1}{2\pi}\int dx[u_s K_s(\partial_x \theta_s)^2+\frac{u_s}{K_s} (\partial_x \phi_s)^2 ] \\
H_a =& \frac{1}{2\pi}\int dx[u_a K_a(\partial_x \theta_a)^2+\frac{u_a}{K_a} (\partial_x \phi_a)^2 ]- \\
     & 2\rho_0 t_\perp\int dx \cos(\sqrt2\theta_a(x))
\end{split}
\end{equation}
Because of the presence of the cosine term the antisymmetric part of the Hamiltonian (the so-called sine-Gordon Hamiltonian) will be massive when $K_a > 1/4$.
As a result the field $\theta_a$ is locked to the minima of the cosine indicating the phase coherence across the two legs of the ladder. The phase being locked, the density fluctuation in the antisymmetric sector have fast decaying correlations instead of the usual power law of the TLL.
The symmetric sector remains massless with power law correlations. A numerical calculation of the TLL parameters for the massless phase can be found in
\cite{crepin_bosonic_ladder_phase_diagram}.

\section{Analytical solutions} \label{sec:analytic}

Let us now investigate the full Hamiltonian (\ref{eq:hamtot}) (or (\ref{eq:hamdiscrete})) to be able to compute the Green's function of the impurity (\ref{eq:greenimp}).

Using (\ref{eq:2eq1c}) the interaction term $H_{\text{coup}}$ with the impurity is expressed, as a function of the bosonized variables as
\begin{multline}
 H_{\text{coup}} = \frac{-U}{\sqrt 2 \pi} \int dx \left(\nabla\phi_a(x)+\nabla \phi_s(x)\right) \rho_{\text{imp}}(x) + \\
         2 U \rho_0 \int dx \cos(\sqrt 2(\phi_a(x)+\phi_s(x)) - 2\pi\rho_0 x) \rho_{\text{imp}}(x)
\end{multline}
where we have retained only the lowest harmonics in the oscillating terms, higher order terms being a priori less relevant.

Since the antisymmetric sector is gapped, with a gap $\Delta_a$ due to the ordering in the field $\theta_a$, the correlation functions involving $\phi_a$ decrease exponentially at large distance or time interval.
We thus a priori need to distinguish the cases for which $U \ll \Delta_a$, for which the interaction with the impurity is not able to create excitations in the antisymmetric sector, from the case $U \gg \Delta_a$. We examine these two cases in turn.

\subsection{$U \ll \Delta_a$}

In this case the coupling with the impurity cannot create excitations in the antisymmetric sector.
The backscattering term containing $\cos(\sqrt 2(\phi_a(x)+\phi_s(x)))$ is thus irrelevant.
This term can potentially generate a $\cos(2\sqrt2 \phi_s(x))$ term by operator product expansion (OPE) but such term would only become relevant for $K_s < 1$ which is outside
the range of allowed TLL parameters for the bosonic problem we consider here.

Thus, as for the case of a single impurity \cite{zvonarev_ferro_cold} the dominant contribution is given by the forward scattering term
\begin{equation}\label{eq:effhamiltonian}
  \frac{-U}{\sqrt 2 \pi} \int dx \nabla \phi_s(x) \rho_{\text{imp}}(x)
\end{equation}
in which we have kept only the part corresponding to the massless sector.

In this regime the impurity behaves in the ladder system as it would be in a single chain with effective parameters $(u_s,K_s)$, but with a renormalized coupling
\begin{equation} \label{eq:renorm}
 U_2 = \frac{U}{\sqrt2}
\end{equation}
Note that for a microscopic interaction $U_1=U_2$ the TLL parameters $(u_s,K_s)$ are also in general not equal to the one of a single chain with the corresponding interaction due to the presence of the irrelevant operators coming from the $t_\perp$ term.

The Green's function of the impurity (\ref{eq:greenimp}) can thus be readily computed by the same techniques used for the single chain case~\cite{zvonarev_ferro_cold,kantian_impurity_DMRG}.
Using a linked cluster expansion (LCE), a re-summed second-order perturbation series applicable at small bath-impurity interactions,
as detailed in Appendix~\ref{ap:LCE}, we find that when $p < \frac{u_s}{2t_{\text{imp}}}$, the difference of energies of the impurity
$\epsilon_p-\epsilon_{p+q}$ intersects the bath dispersion $u_s|q|$ only at $q=0$. As shown in Appendix~\ref{ap:LCE} this leads
to a power law decay of the Green's function (\ref{eq:greenimp})
\begin{equation} \label{eq:2eq1}
|G(p,t)| = e^{-\frac{K_s U^2}{4 \pi^2 u_s^2}(1+\frac{12 t_{\text{imp}}^2 p^2}{u_s^2})\log(t)}
\end{equation}
This case will be discussed in details in Sec.~\ref{sec:zeromoment}.

On the contrary for $p>\frac{u_s}{2t_{\text{imp}}}$, $\epsilon_p-\epsilon_{p+q}$ intersects the bath dispersion $u_s|q|$  at non-zero $q's$. In that case the Green's function decay exponentially, and enter into the so called quasiparticle (QP) regime \cite{kantian_impurity_DMRG},
where $\epsilon_p=-2t_{\text{imp}}\cos(p)$. This case will be discussed in Sec.~\ref{sec:finitemoment}.\\
Dynamical structure factor $S(k,\omega)$ of one of leg of the ladder is given by
\begin{equation} \label{eq:structure_factor}
\begin{split}
S(x,\omega) &= \langle GS_b|\rho_1(x)\frac{1}{\omega+i \eta-H_{\text{lad}}}\rho_1(0)| GS_b\rangle\\
S(k,\omega)&=\int dx S(x,\omega)\exp(-ikx)
\end{split}
\end{equation}
Where $\eta\rightarrow 0^+$, $\rho_1(x)$ is the density operator in leg 1 of the ladder at position $x$.\\
To go beyond the bosonized evaluation of the structure factor of the bath used in Appendix~\ref{ap:LCE},
we also show in Fig.~\ref{fig:Structure_factor}
\begin{figure}
\begin{center}
  \includegraphics [scale=0.5]
{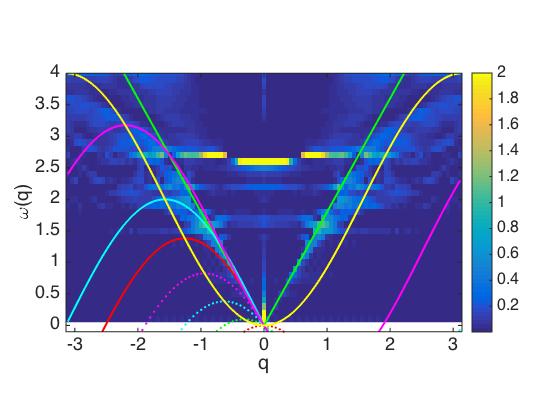}
\end{center}
\caption{\label{fig:Structure_factor} (color online)
Dynamical structure factor leg $1$ of the two-leg hardcore boson ladder at one third filling (shaded area).
Superimposed colored curves show the change in the impurities kinetic energy when emitting an excitation with momentum $q$ into the bath, $\delta\epsilon(q)=\epsilon_\p-\epsilon_{p+q}$, for different values of $p$. In ascending order by maximal value these are: red ($p=0$), green ($0.1\pi$), cyan ($0.2\pi$), magenta ($0.3\pi$), red ($0.4\pi$), cyan ($0.5\pi$), magenta ($0.7\pi$) and yellow ($\pi$) . Dotted lines ($p=0, 0.1\pi, 0.2\pi$) denote where  $\delta\epsilon(q)$ intersects with areas of finite weight of the dynamical structure factor only at $q=0$. Solid lines and dotted line at $p=0.3\pi$ denote the higher $p$-values for which it intersects also at non-zero $q$'s. This change in regime causes the impurity to go from subdiffusive to quasiparticle behavior (see text).}
\end{figure}
the imaginary part of dynamical structure factor define in Eq. (\ref{fig:Structure_factor}) of one of leg of a hard core boson ladder computed using DMRG at $t_\perp=t_b=1, \chi=600$ and zero temperature, together with $\delta\epsilon(q)=\epsilon_p-\epsilon_{p+q}$, for different $p$ values at  $t_{\text{imp}}=1$.
We find that for small $q$ the dispersion of bath is well fitted with the linear dispersion used in the TLL representation
$u_s|q|$ (green line, $u_s=1.8$). We find, in agreement with the LCE, that for $p=0,0.1\pi, 0.2\pi$, $\delta\epsilon(q)$ intersects the dynamical structure factor only at $q=0$, while for large $p=0.3\pi, 0.4\pi, 0.5\pi, 0.7\pi, \pi$ , $\delta\epsilon(q)$  intersects the dynamical structure factor  at $q\neq0$. The resulting change of regime for the impurity will be discussed in more details
in Sec.~\ref{sec:finitemoment} and Fig.~\ref{fig:decay}.

\subsection{$U \gg \Delta_a$}

We now consider the more complicated opposite case for which the interaction with the impurity, which couples to the field $\phi_1$,  can potentially induce excitations across the antisymmetric gap for the field $\theta_1$. In order to deal with this case we essentially follow the method introduced by Lamacraft for the case of the single chain \cite{Lamacraft_mobile_impurity_in_one_dimension}.

We first make a transformation to a frame of reference which moves with the  impurity, and this can be imposed by an unitary transformation $U_X=e^{i(P_1+P_2)X}$, where $P_1$ , $P_2$ and $X$ are total momentum operator of the first, second leg of the ladder  and position operator of the impurity respectively. In bosonized language $P_1=\int dx \nabla\theta_1\nabla\phi_1$, $P_2=\int dx  \nabla\theta_2\nabla\phi_2$.

By using (\ref{eq:hamtot},\ref{eq:imp_density}), the effective Hamiltonian in the new frame of reference is given by
\begin{eqnarray}
H_{\text{eff}}&=&U_X H U_X^\dagger  \nonumber \\
&=&\frac{(P-P_1-P_2)^2}{2 M}+H_{\text{lad}}+U\rho_1(0)
\label{eq:ham_eff}
\end{eqnarray}
Where $\rho_1(0)$ is the density of leg $1$ at $x=0$. In the new effective Hamiltonian the position operator $X$ has disappeared and $P$ is thus a conserved quantity.
As $U\rightarrow \infty$, $U$ is replaced by an effective forward scattering potential $U_\phi$ (roughly the phase shift corresponding
to the potential $U$) \cite{Lamacraft_mobile_impurity_in_one_dimension} and $\rho_1=-\frac{\nabla \phi_s+\nabla \phi_a}{\sqrt 2\pi}$.
The forward scattering can thus be absorbed in the quadratic term by creating a discontinuity in the fields $\phi$ of the form
\begin{equation}
\begin{split}
 \phi_s|_{0^+}^{0^-} &= \frac{K_sU_\phi}{\sqrt2 u_s} \\
 \phi_a|_{0^+}^{0^-} &=\frac{K_aU_\phi}{\sqrt2 u_a}
\end{split}
\end{equation}
The dominant term in the total current is given by
\begin{equation}
 P_1 + P_2 = \int dx \rho_0 \sqrt2 \nabla\theta_s(x)
\end{equation}
where in the above formula the origin must be excluded if the field has a discontinuity.
For $P=0$ the minimization of $H$ imposes that the field $\theta_s$ remains continuous at $x=0$.
These two set of conditions for the fields $\phi$ and $\theta$ can be imposed on otherwise continuous
fields by the unitary transformation
\begin{equation}
 U_{P=0}=\exp(i \theta_s(0)\frac{K_sU_\phi}{\sqrt2 \pi u_s})\exp(i \theta_a(0)\frac{K_aU_\phi}{\sqrt2 \pi u_a})
\end{equation}
The impurity Green's function at zero momentum is thus given by
\begin{equation}\label{eq:2eq31}
\begin{split}
|G(0,t)|&= \langle U^\dagger_{P=0,t}  U_{P=0,0}\rangle  \\
&=\langle  \exp(-i \theta_s(t)\frac{ K_sU_\phi}{\sqrt2\pi u_s}) \exp(i \theta_s(0)\frac{K_sU_\phi}{\sqrt2\pi u_s})\rangle \\ &=  \langle   \exp(-i \theta_a(t)\frac{K_aU_\phi}{\sqrt2\pi u_a}) \exp(i \theta_a(0)\frac{K_aU_\phi}{\sqrt2\pi u_a})\rangle  \\
 &= |t|^{-K_s/4 (\frac{U_\phi}{\pi u_s})^2}
\end{split}
\end{equation}

To compute the above correlation function corresponding to field $\theta_a$  we have expanded $\cos(\sqrt2 \theta_a)$ upto second order, and exploited that the correlation function of the anti-symmetric mode saturates to a finite value for  time greater than the inverse of  gap in the anti-symmetric sector. The symmetric mode decays as a power law with an exponent $\frac{K_s}{4} (\frac{U_\phi}{\pi u_s})^2$. For hard core bosons in one dimension  $\frac{U_\phi}{\pi u_s}=1$, and thus the overall exponent in (\ref{eq:2eq31}) is $K_s/4$.

\section{Numerical solution} \label{sec:numerics}

\subsection{Method}

In order to obtain the Green's function of the impurity for the ladder problem quantitatively, we compute a numerical solution of the
lattice model defined in Sec.~\ref{sec:model}, by a method analogous to the one used for the single chain~\cite{kantian_impurity_DMRG}.
We use DMRG to compute the ground state of the bath and time-dependent DMRG (t-DMRG) to compute the Green's function of the impurity.

For t-DMRG we use a supercell approach to map the three species of bosons $(A,B,C)$ into a one dimensional chain.
Leg~1 and leg~2 of the ladder are represented by species $A$ and $B$ respectively and the impurity is represented by $C$. The total number of quantum particles in leg $A$ and
leg $B$ is conserved, and the total number of particles in species $C$ is conserved separately and equal to $1$. In the supercell approach, the local Hilbert space is of size $2 \times 2 \times 2 = 8$
for the case of hard core bosons and of $3 \times 3 \times 2 = 18$ for soft core bosons if the maximum allowed
occupancy per species $A,B$ is $2$. We thus restrict ourselves to relatively large repulsions (of the order of $U_1 = U_2 = 10$ for softcore bosons and $U_1 = U_2 = \infty$ for hardcore ones)
in each leg for the bath so that the restriction of the Hilbert space is not a serious limitation.
The interaction between the bath and the impurity can however take any value.

We consider a system for which the density in each leg is $\rho_0 = 1/3$, to avoid the possibility of entering to a Mott insulating state in the ladder in which the symmetric sector would be gapped as well. Transverse and impurity hopping are taken of equal value, $t_\perp = t_{\text{imp}}=1$. We fix the size of the system to $L=101$ sites per leg.

In t-DMRG, the singular value decomposition of a matrix of order of $(d \chi)\times (d\chi)$ is needed~\cite{Schollwock_density_matrix_renormalization_group} for a sweep through a
bond between two lattice sites, where $d$ is local Hilbert space dimension and $\chi$ is bond dimension which encodes the amount of entanglement in the system.
At each time step, $(L-1)$ such operations are performed.
As the local Hilbert space can become large for soft-core bosons, we are limited to moderate values of $\chi$ to maintain reasonable computational times.
We further use bond dimensions $\chi=300, 500, 400, 600$ for hard core boson and $\chi=400$ for soft core boson.
Further details are provided in Appendix~\ref{ap:dmrg}.

The ground state of the bath $|GS_b\rangle$ is computed by using DMRG. We then add
an impurity in the center of leg~1, at time $t=0$
\begin{equation}
 |\psi(t=0) \rangle = d^\dagger_{\frac{L+1}2} |GS_b\rangle
\end{equation}
We then evolve the state $|\psi(t=0)\rangle$  as a function of time($t$) with the full Hamiltonian (\ref{eq:hamdiscrete}) by using t-DMRG and compute
\begin{equation}
 |\psi(t)\rangle = e^{-i H t} d^\dagger_{\frac{L+1}2} |GS_b\rangle
\end{equation}
We then take the overlap with the state
\begin{equation}
 d^\dagger_{\frac{L+1}2-x} |GS_b\rangle,
 \end{equation}
which yields the sought-after impurity Green's function in time and space, up to a phase factor of $e^{i E_{GS_b} t}$,
where $E_{GS_b}$ is ground state energy of the bath. A Fourier transformation then yields the Green's function in time and momentum.

\subsection{Results at zero momentum} \label{sec:zeromoment}

We examine first the Green's function of the impurity $G(p,t)$ for the case of zero momentum $p=0$. For a single chain this limit is known to lead to a power law decay of the
Green's function~\cite{zvonarev_ferro_cold,Lamacraft_mobile_impurity_in_one_dimension,kantian_impurity_DMRG}.

\begin{figure}
\begin{center}
  \includegraphics [ scale=0.25]
 {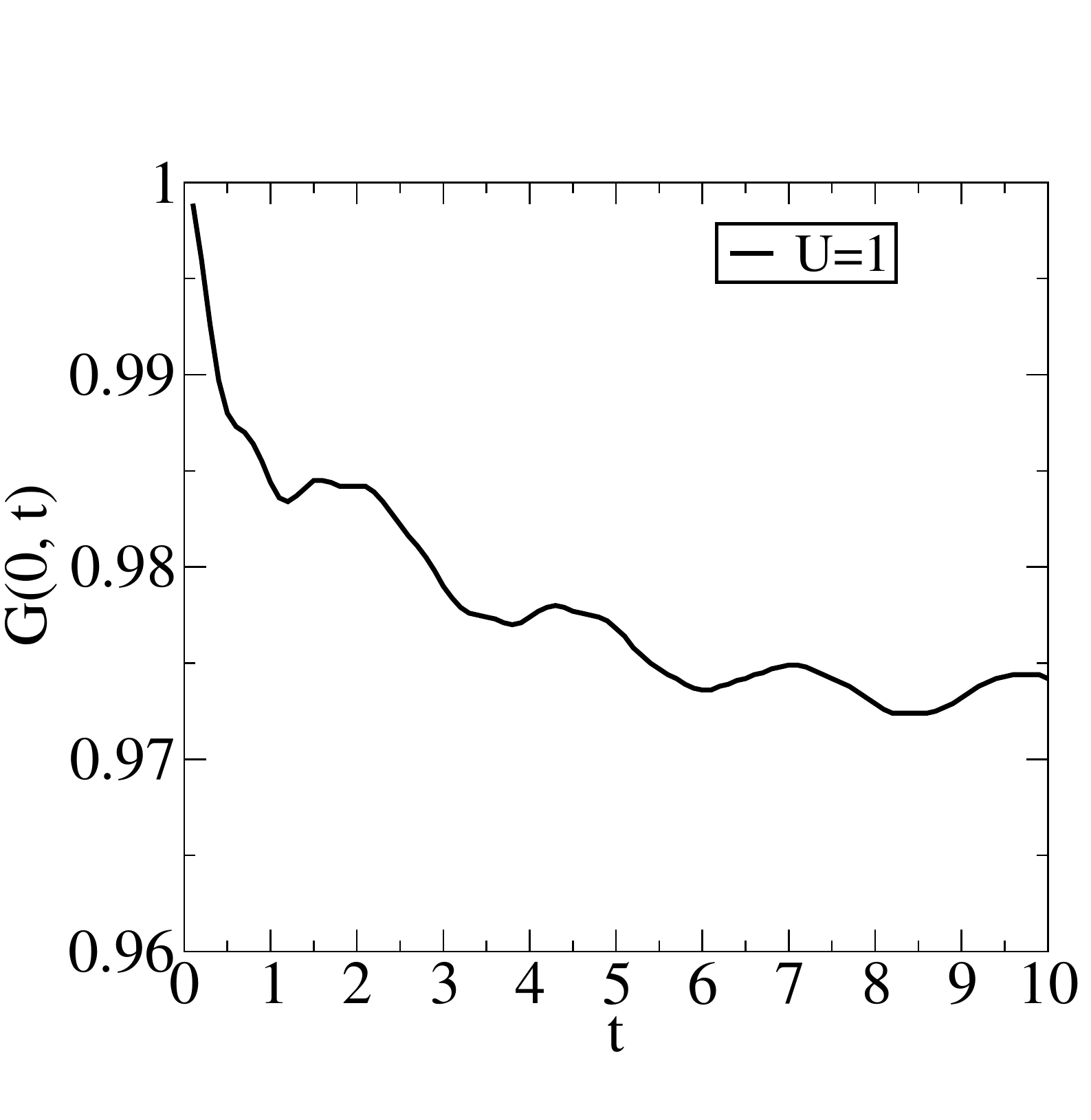}
  \includegraphics [ scale=0.25]{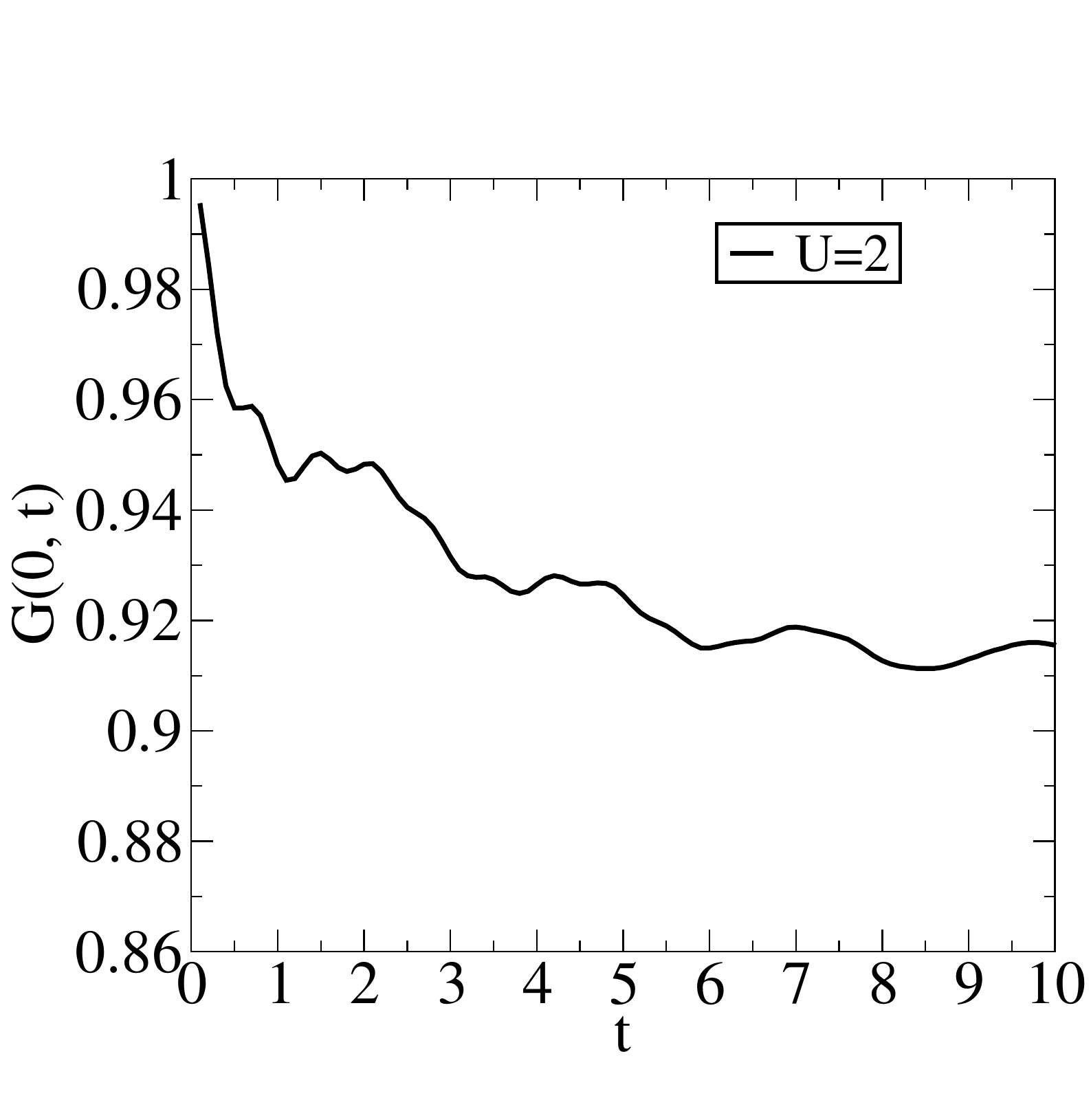}
   \includegraphics [ scale=0.25]
 {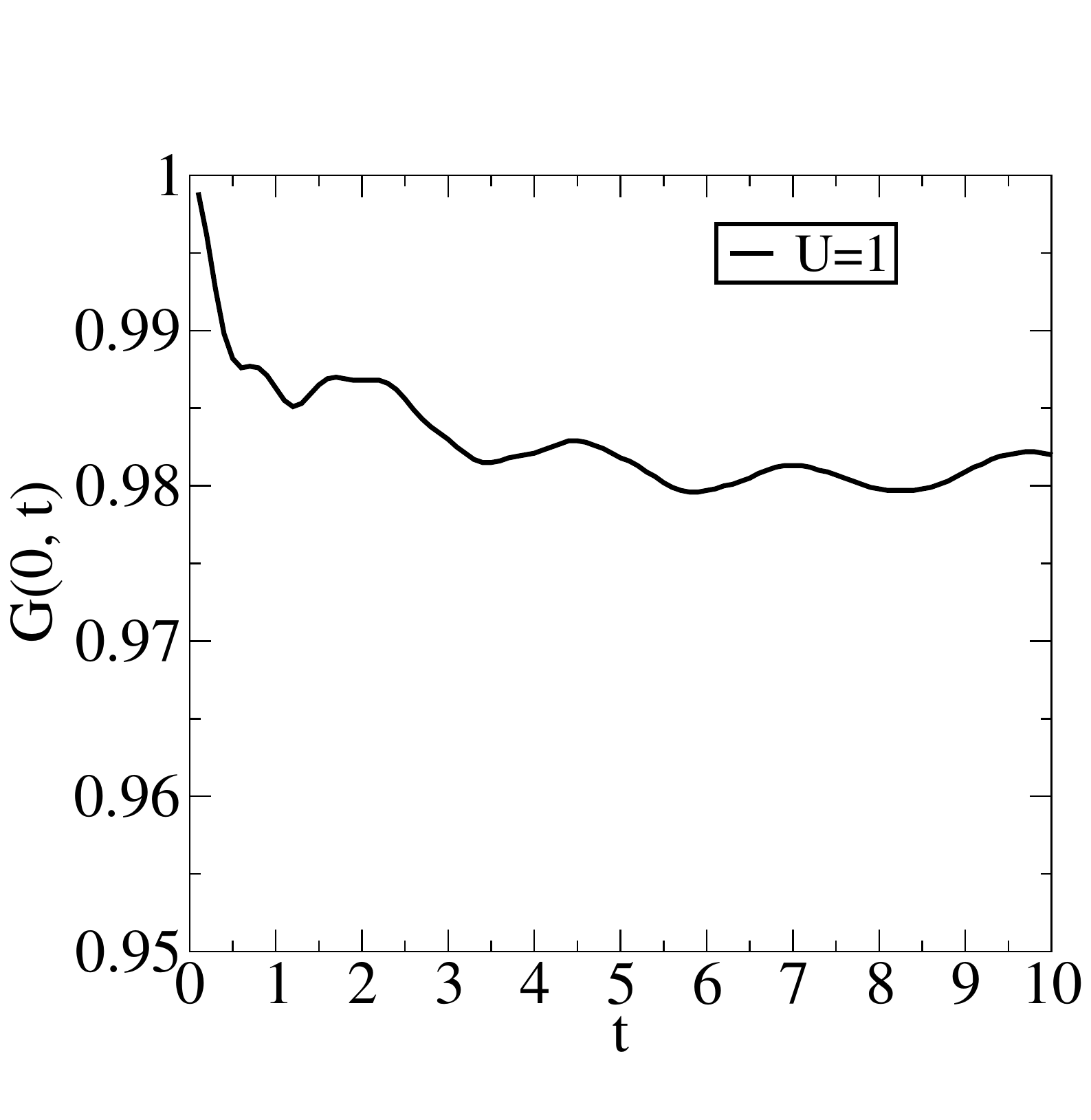}
  \includegraphics [ scale=0.25]{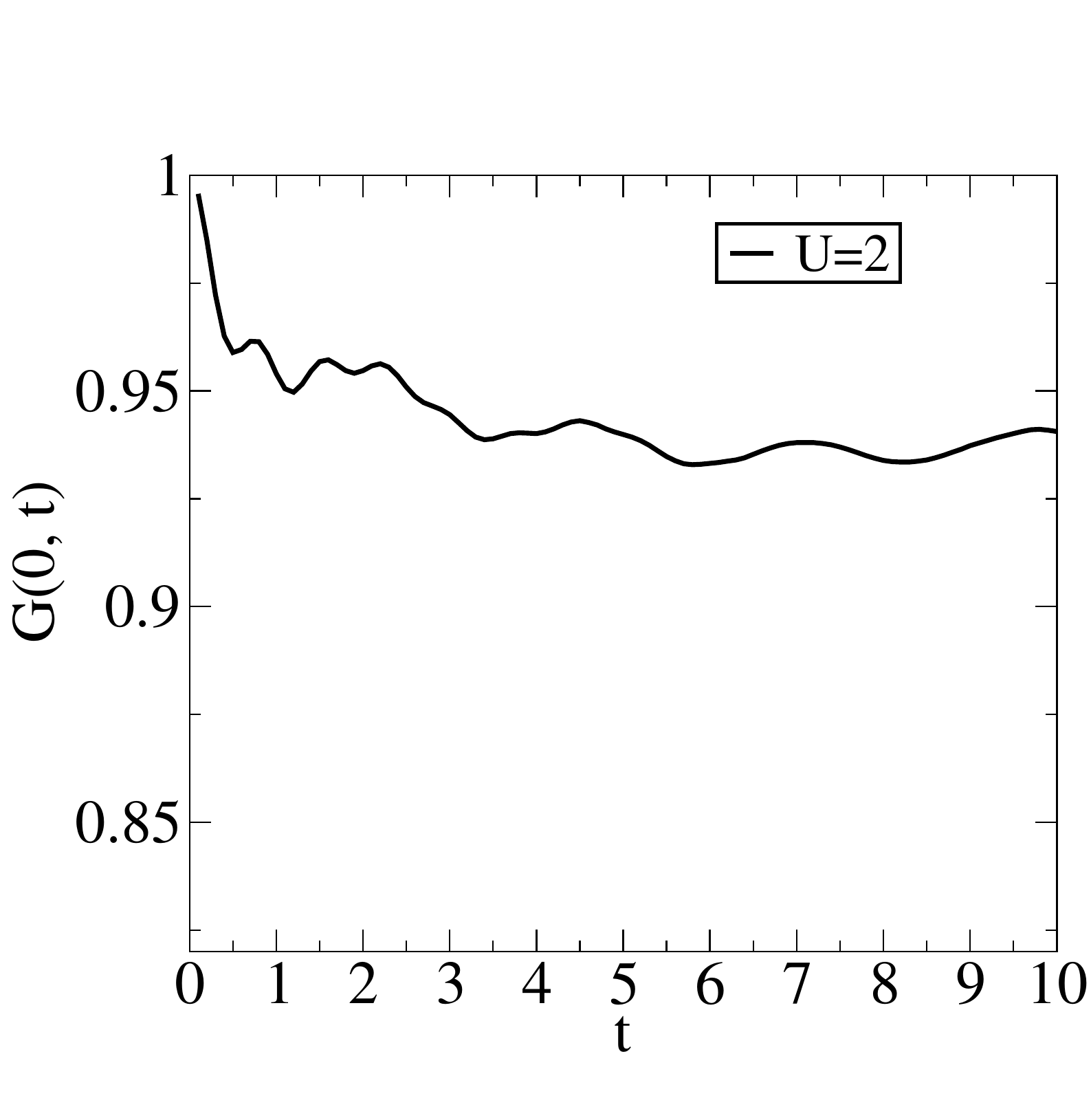}
\end{center}
\caption{\label{fig:2fig2}
(color online) Modulus of the Green's function of the impurity (see text) $|G(p=0,t)|$ at zero momentum of the
impurity.
Parameters for the interchain hopping , impurity hopping, and impurity-bath interaction in the ladder
are respectively $t_\perp=1$, $t_{\text{imp}} = 1$. Left (resp. right) column corresponds to $U=1$ (resp. $U=2$).
The upper panel shows $|G(0,t)|$ of softcore bosons with $U_1=U_2=10$ and the lower panel hardcore bosons for $\chi=400$.}
\end{figure}

Examples of the decay of $|G(0,t)|$ are shown in Fig.~\ref{fig:2fig2}, both for softcore bosons and hard core bosons.
The numerical data shows a decay of the correlation function with time. Such a decay is expected from the general arguments of Sec.~\ref{sec:analytic}.
As discussed in Appendix~\ref{ap:dmrg} the bond dimension controls the maximum time at which the
decay of the correlation can be computed reliably with the t-DMRG procedure. In our case, time of the order $t\sim 7$ for soft-core and $t\sim 7$ for hard-core bosons present the limit
for reliable data.

\subsubsection{Interaction dependence}

\begin{figure}
\begin{center}
\includegraphics[scale=0.4]{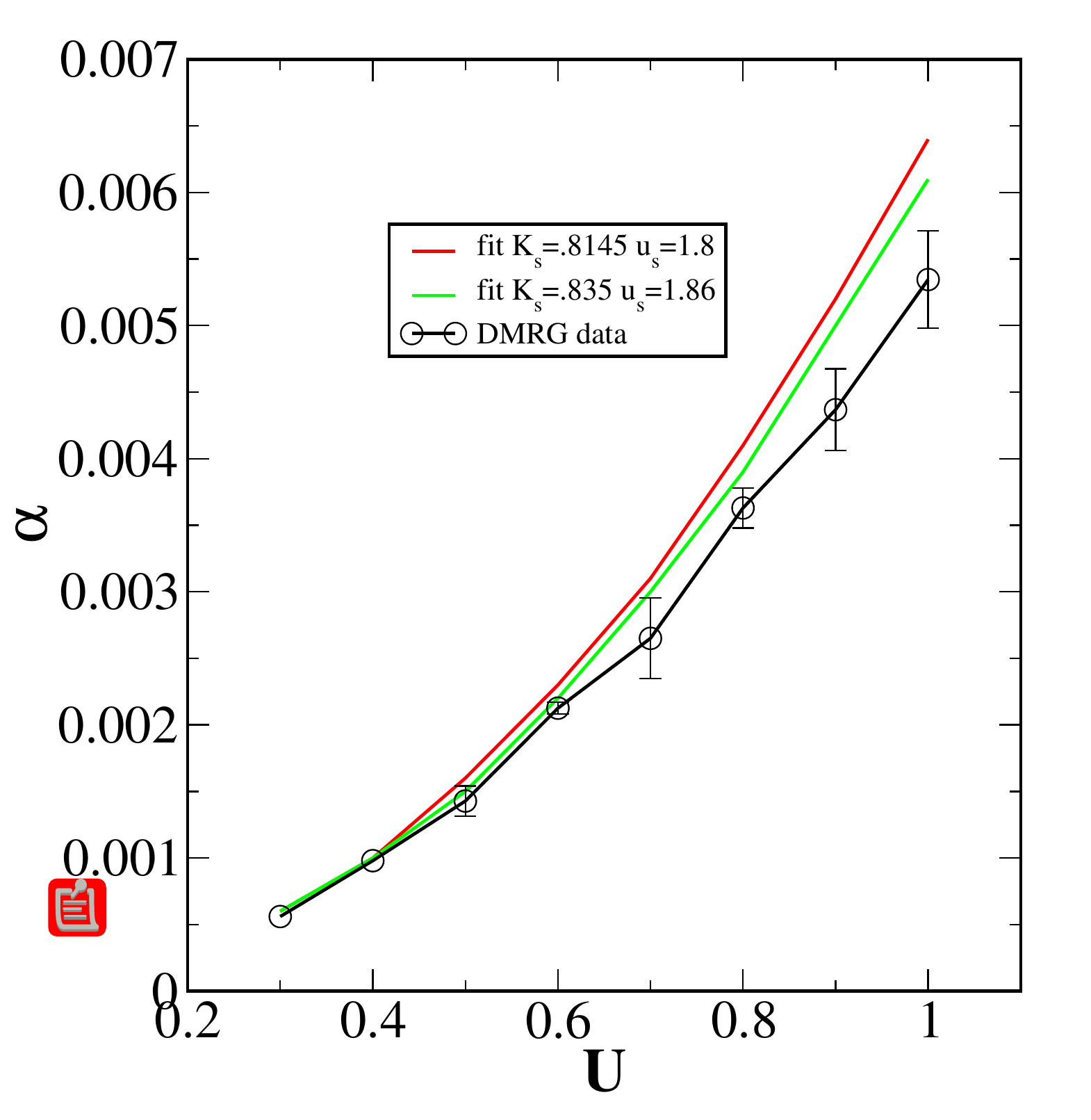}
\end{center}
\caption{\label{fig:2fig3}(color online)
Green's function exponent at $p=0$ for hard core bosons as function of small $U$ at $t_\perp=t_{\text{imp}}=1$.
The black curve is the exponent extracted from correlation functions computed from t-DMRG results with $\chi=600$. The green and red curves are the LCE exponent extracted from (\ref{eq:2eq1}),
$\alpha_{LCE} = \frac{K_sU^2}{4 \pi^2 u_s^2}$, for the pair of values ($K_s=.835$, $u_s=1.86$), ($K_s=.8145$, $u_s=1.8$) respectively. These values correspond well to the TLL parameters for a ladder of hard core bosons (see text).}
\end{figure}

To analyze the data we use the analytic estimates of Sec.~\ref{sec:analytic} which suggest a power law decay of the Green's function.
\begin{equation}
 |G(p=0,t)| \propto \left(\frac{1}{t}\right)^{\alpha}
\end{equation}
We fit the numerical data as detailed in Appendix~\ref{ap:dmrg},
which confirms the power law decay of the correlations and allows extracting the exponent $\alpha$. This exponent is shown in Fig.~\ref{fig:2fig3} for the case of hard core bosons.

The good agreement between the numerical exponent of Fig.~\ref{fig:2fig3} with the LCE-formula in (\ref{eq:2eq1}) confirms the analytic prediction of Sec.~\ref{sec:analytic} that
one can indeed view the ladder with a small impurity-bath interaction $U$ as a single TLL with an effective $K_s$ and $u_s$ but with an effective interaction $U/\sqrt2$.
The simple decoupling of the Hamiltonian into symmetric and antisymmetric sector of (\ref{eq:2eq7}) would naively suggest
that $(K_s,u_s) = (K,v)$ of a single chain,
but for large $U_1=U_2$ interactions and sizeable $t_\perp$, irrelevant operators can lead to a sizeable renormalization of the parameters. So in general the parameters $(K_s,u_s)$ for the ladder are not identical to the ones of a single chain with the same
interaction, leading also to a modification of the exponent.
This is in particular the case for hard core bosons, for which the single chain parameter is $K=1$~\cite{Giamarchi_Bosonization}, while for the ladder~\cite{crepin_bosonic_ladder_phase_diagram} one has $K_s < 1$
Our calculations yield $K_s = 0.8145$
and $u_s=1.8$. Given the accuracy of determination of the TLL parameters this compares well with the values obtained in  Ref.~\onlinecite{crepin_bosonic_ladder_phase_diagram} which are $K_s=0.835$ and $u_s=1.86$  (c.f. also Fig.~19 of~\cite{crepin_bosonic_ladder_phase_diagram}).
We note that for the single chain a value of $K<1$ would have meant that the backscattering on the impurity terms could become relevant.
For the ladder case however , because the antisymmetric sector is gapped, such terms remain irrelevant even for $K_s < 1$.

\begin{figure}
\begin{center}
  \includegraphics [scale=0.5] {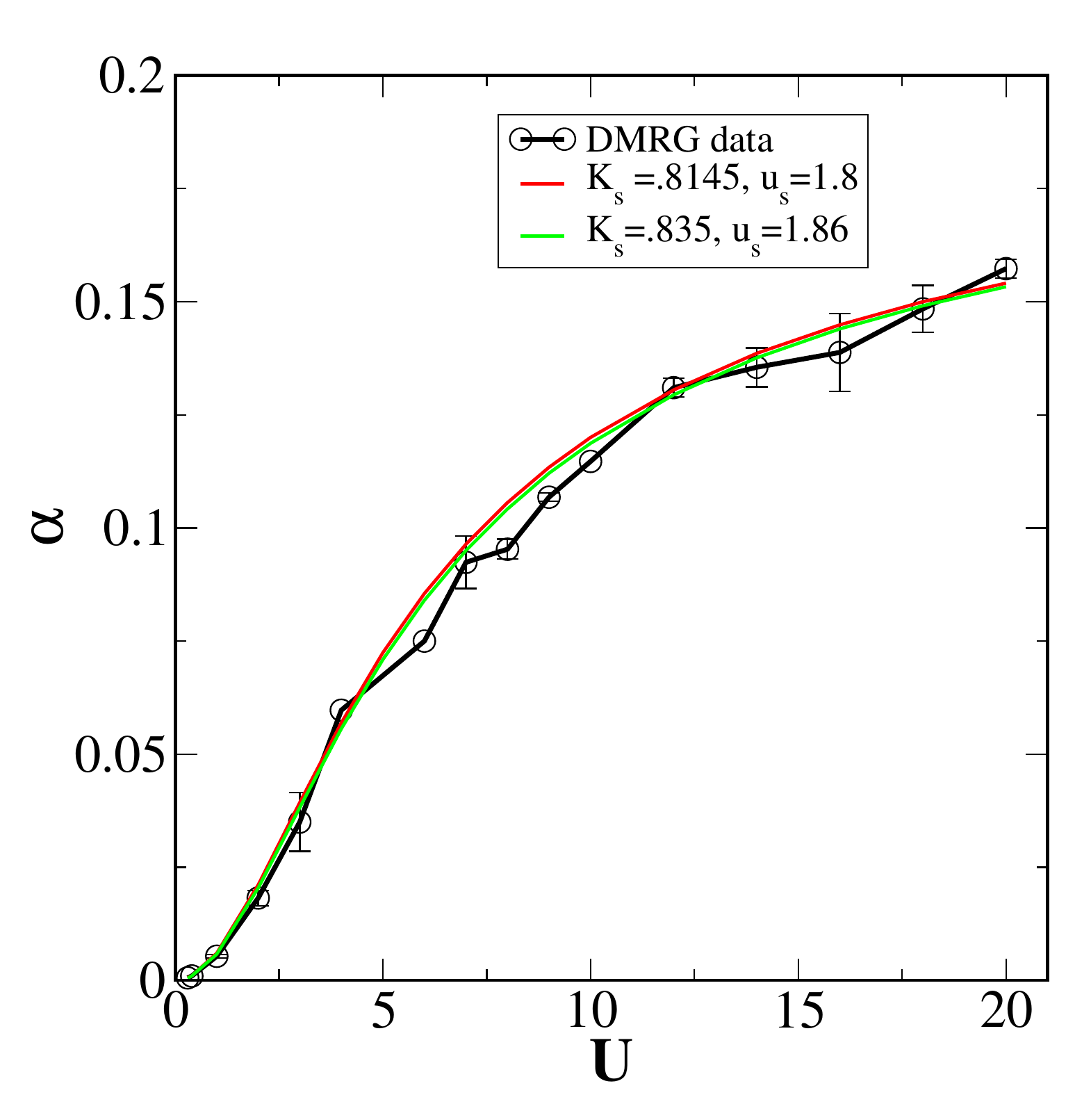}
\end{center}
\caption{\label{fig:interaction}(color online) Exponent $\alpha$ controlling the power-law decay of $|G(p=0,t)|$ as a function of the interaction between the bath and the impurity $U$. The bath has
$t_\perp=t_{\text{imp}}=1$ and is made of hard core bosons at $1/3$ filling for each leg. The black circle are the numerical data.
The red line is a fit to (\ref{eq:empform}) (see text) with $f=0.39$. The good agreement between the data and the formula
(\ref{eq:empform}) shows that, as for a single chain, this formula correctly describes the behavior of the impurity in the ladder for a wide range of interactions.}
\end{figure}

For larger values of the impurity-bath interaction $U$ one cannot rely on the LCE expression anymore. The numerically computed exponent is shown in Fig.~\ref{fig:interaction}.

The quadratic growth of the exponent with interaction of Fig.~\ref{fig:2fig3} is replaced by a more complex behavior and a saturation of the exponent at large $U$.
For the single chain the interaction dependence of the exponent could be captured by an analytic expression \cite{kantian_impurity_DMRG}. In order to adapt this
expression to the case of ladder we use the modified interaction $U/\sqrt{2}$ suggested by the bosonization formula of Sec.~\ref{sec:analytic} and the TLL parameters of the ladder which leads to
\begin{equation} \label{eq:empform}
 \alpha = \frac{2 f}{\pi^2} \left(\arctan\left[\frac{2 \sqrt{2 f u_s^2}}{U\sqrt K_s}\right]- \frac{\pi}2 \right)^2
\end{equation}
Here, $f=2\alpha[U\rightarrow\infty]$, and we use the independently calculated exponent at $U=\infty$ (see next section) to obtain $f=0.39$ for plotting the analytical curves in Fig.~\ref{fig:interaction}. We stress that the excellent agreement between DMRG data and analytical prediction eq.~(\ref{eq:empform}) shown in Fig.~\ref{fig:interaction} is thus without fitting parameters, unlike that in Ref.~\cite{kantian_impurity_DMRG}, which had been lacking an independent way of obtaining $f$ quantitatively.

\subsubsection{Hard core bath-impurity repulsion}

\begin{figure}
\begin{center}
\includegraphics[scale=0.5]
 {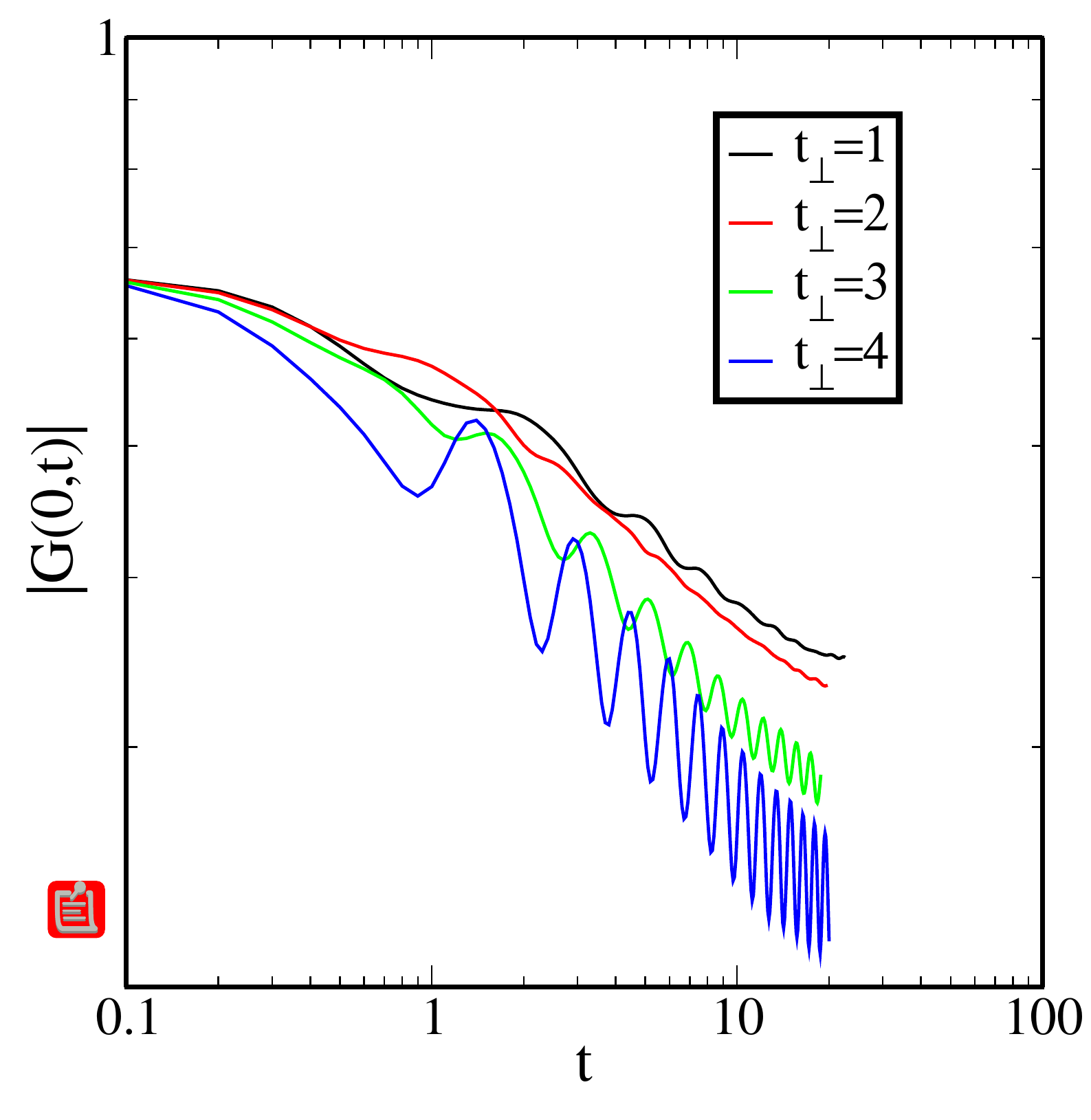}
\caption{\label{fig:2fig4} (color online)  Green's function of  the impurity at zero momentum on a log-log scale for the ladder of hard core bosons and for infinite repulsion between impurity and bath particles. $t_\perp = 1,2,3,4$ and $t_b = t_{\text{imp}} = 1$. The t-DMRG has been performed for $\chi=600$. The powerlaw decay of the correlation is present for all measured values of $t_\perp$.}
\end{center}
\end{figure}

\begin{figure}
\begin{center}
 \includegraphics[scale=0.5]
  {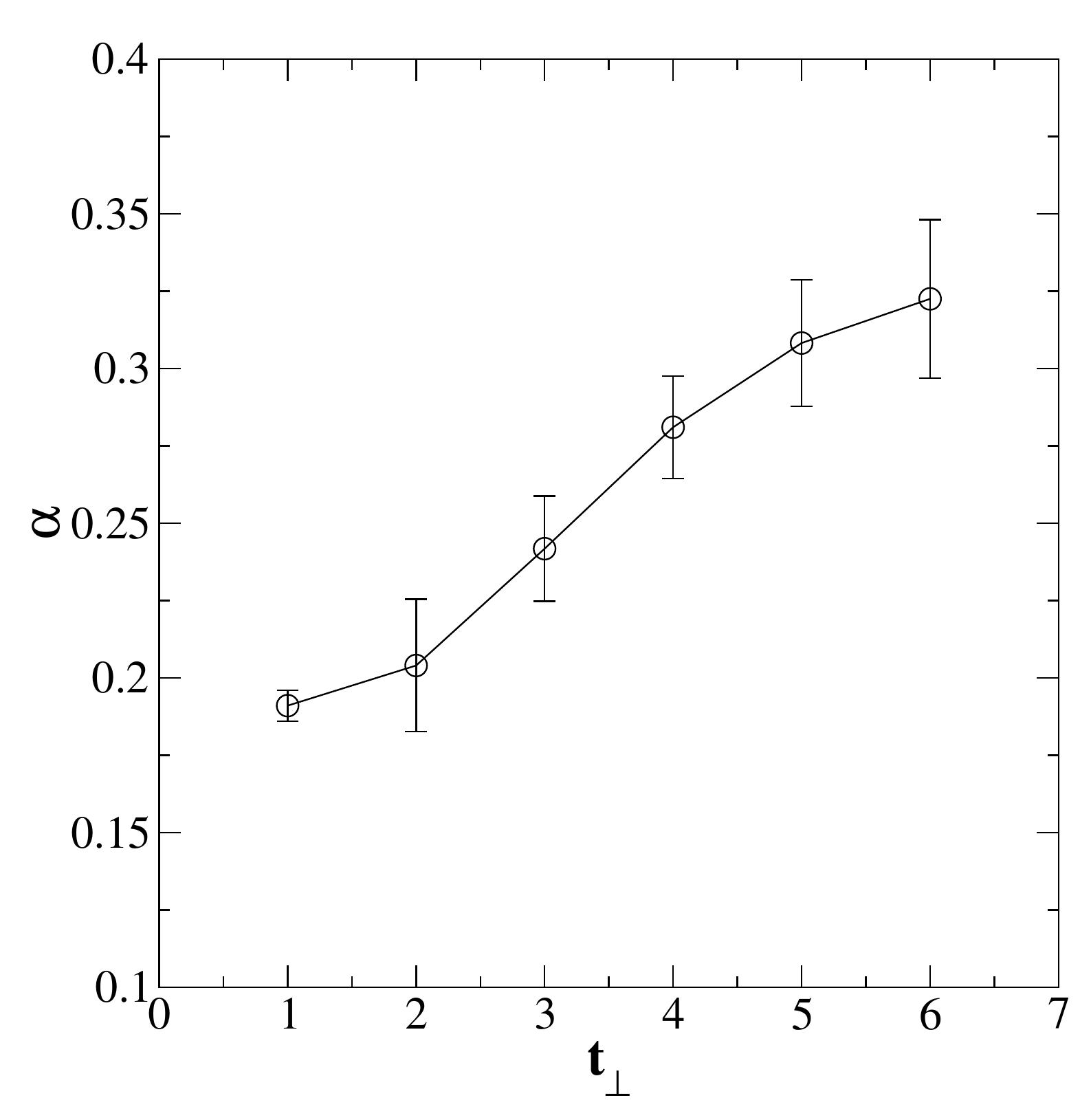}
\end{center}
\caption{\label{fig:2fig5} (color online) Exponent of the Green's function of an impurity with a hard core repulsion with a bath of hard core bosons at filling $1/3$ as a function of $t_\perp$ at $t_{\text{imp}} = 1$, $U\rightarrow\infty$ and $\text{p=0}$. Circles are the numerical data for  $\chi=600$ and the line is a guide to the eyes.}
\end{figure}

Let us now turn to the case for which the repulsion between the impurity and particles of the bath is very large $U \to \infty$.
Various decays of the Green's function are shown in Fig.~\ref{fig:2fig4}.
As is clear from the numerical data the powerlaw decay is still present in this limit and persists for all the measured values of $t_\perp$ in the ladder.
A fit of the numerical data provides access to the exponent as a function of interchain hopping as shown in Fig.~\ref{fig:2fig5}.
We first note that the value of the exponent for $U=\infty$ at $t_\perp = 1$, $\alpha = .195$ is in good agreement with the one given by the formula (\ref{eq:empform}).
Larger values of $t_\perp$ show a marked increase of the exponent as is obvious from Fig.~\ref{fig:2fig3}. This trend of the exponent with $t_\perp$ is surprising and is a priori not
compatible with the simple extrapolation of the formula  $\frac{K_s U^2}{4 \pi^2 u_s^2}$ in which the interaction $U$ would be replaced by its phase shift as for the single
chain \cite{Lamacraft_mobile_impurity_in_one_dimension} where $U \to U_\phi = v \pi$.
In such a limit the exponent would be proportional to $K_s$ which decreases when $t_\perp$ increases
at variance with the numerical data. This shows that there is a dependence on $t_\perp$ of the exponent besides the one hidden
in the $t_\perp$ dependence of the TLL parameters.

\subsection{Momentum dependence of the exponent} \label{sec:finitemoment}

We now turn to the momentum dependence of the Green's function, which, as for the single chain, is much more difficult to obtain
analytically.

\begin{figure}
\begin{center}
  \includegraphics [ scale=0.4]
 {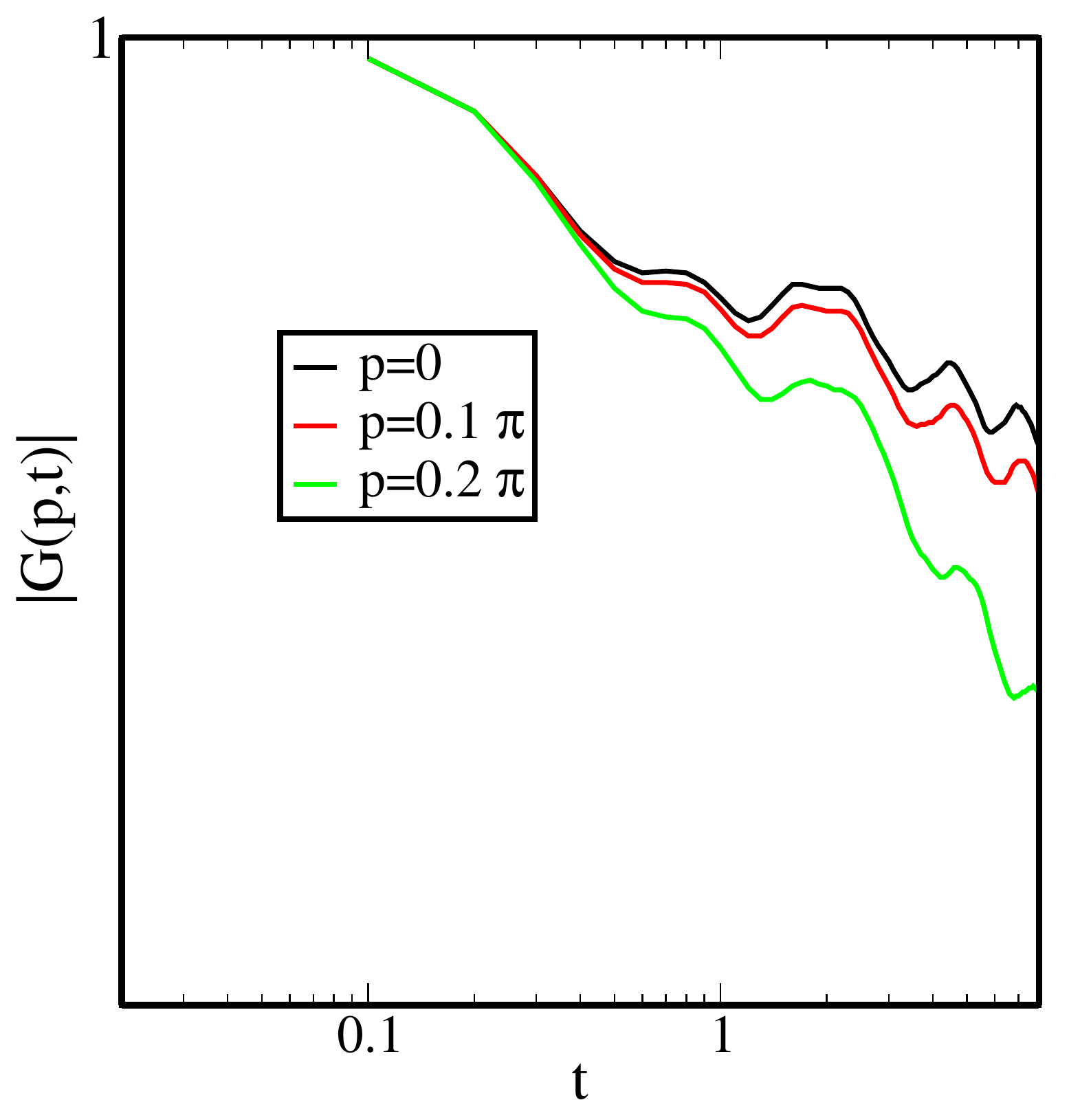}
  \includegraphics [ scale=0.4]{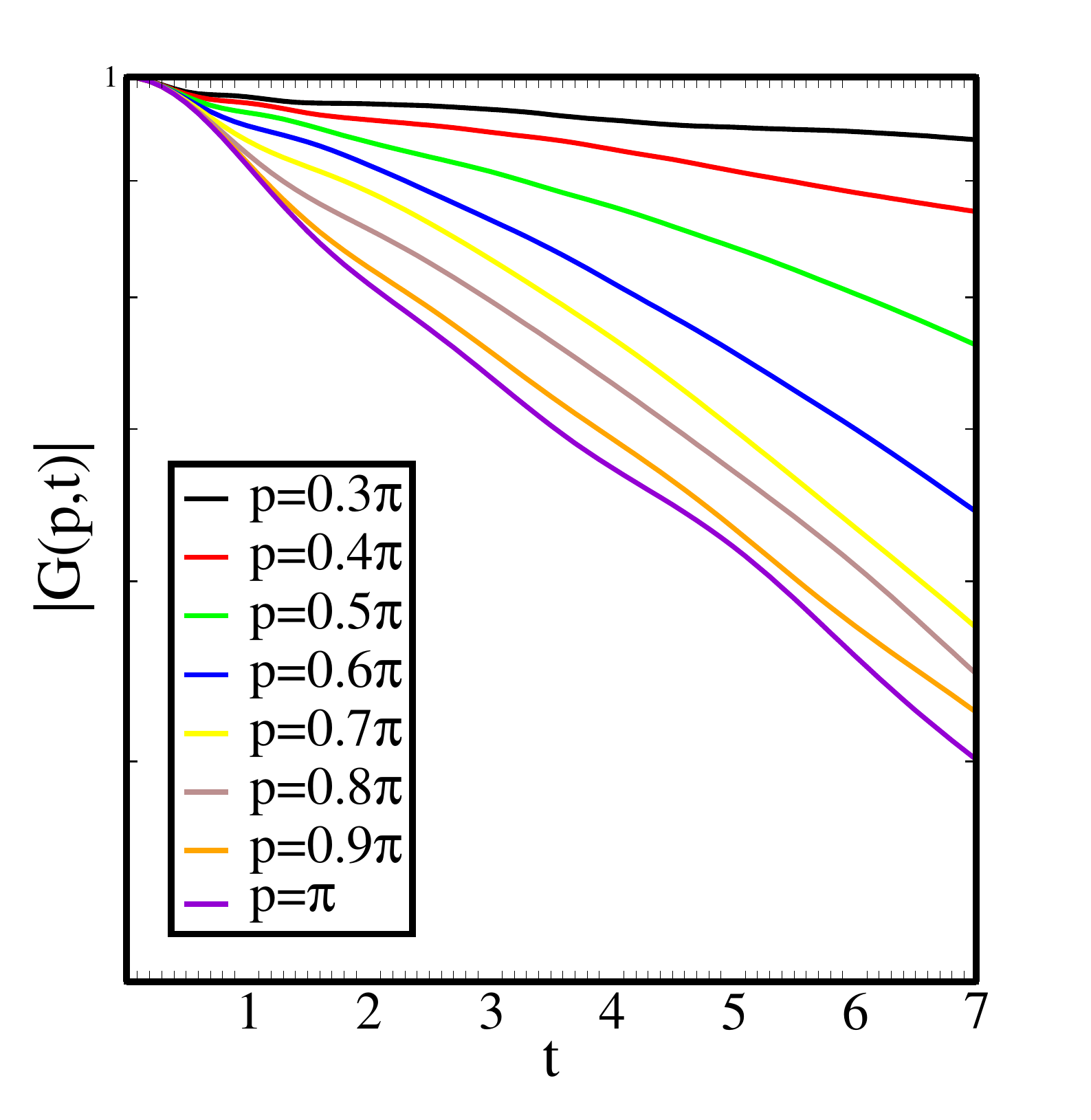}
\end{center}
\caption{\label{fig:decay} (color online)  Green's function of the impurity in a ladder of hard core bosons,  $t_{\text{imp}}=1$,  and transverse hopping($t_\perp=1$), $\text{U}=1$, for different momenta and  $\chi=600$.
Upper panel: Green's function on log-log scale. The momenta are $p=0, 0.1\pi, 0.2\pi$. On a log-log scale the Green's functions show a linear behavior which reflects the power law decay.
Lower Panel: Green's function on semi log plot. Momenta are $p=0.3\pi, 0.4\pi, 0.5\pi, 0.6\pi, 0.7\pi, 0.8\pi, 0.9\pi, \pi$.
On the semi log scale the Green's function show a linear behavior which reflects the exponential decay.}
\end{figure}

We show in Fig.~\ref{fig:decay} the decay of the Green's function for various momenta $p$ of the impurity.
The top panel, for small momentum of the impurity, shows a power law decay of the Green's function, similar to the one for zero momentum, albeit with a renormalized exponent. However, above a momentum threshold around $0.3 \pi$ the numerical data no longer exhibits a power-law decay, and is fitted better with an exponential decay, as shown in the lower panel of the figure. This is similar to the behavior observed for an impurity coupled to a single chain
\cite{kantian_impurity_DMRG}.

\begin{figure}
\begin{center}
  \includegraphics [ scale=0.4]
 {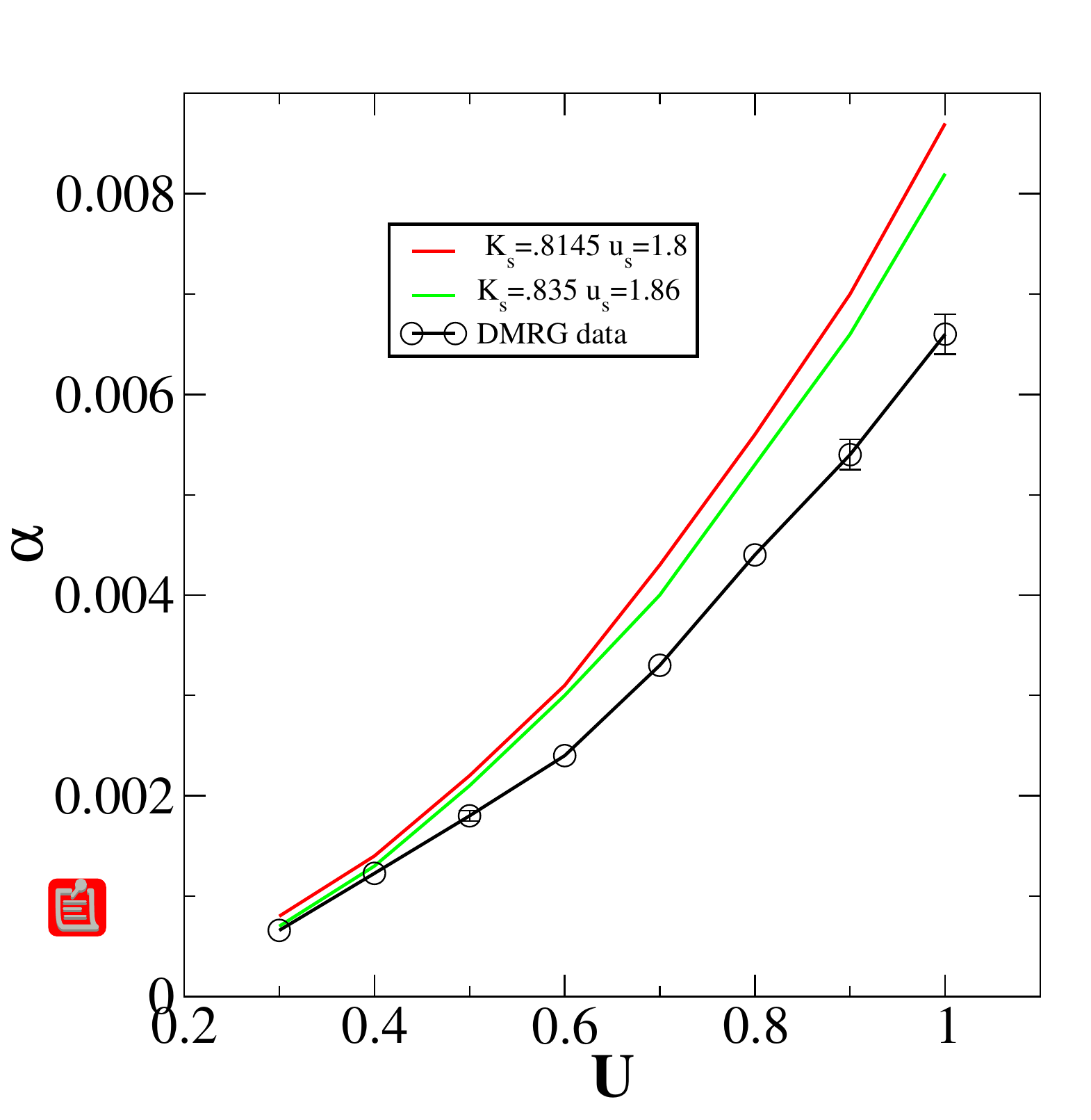}
 \includegraphics [ scale=0.4]
 {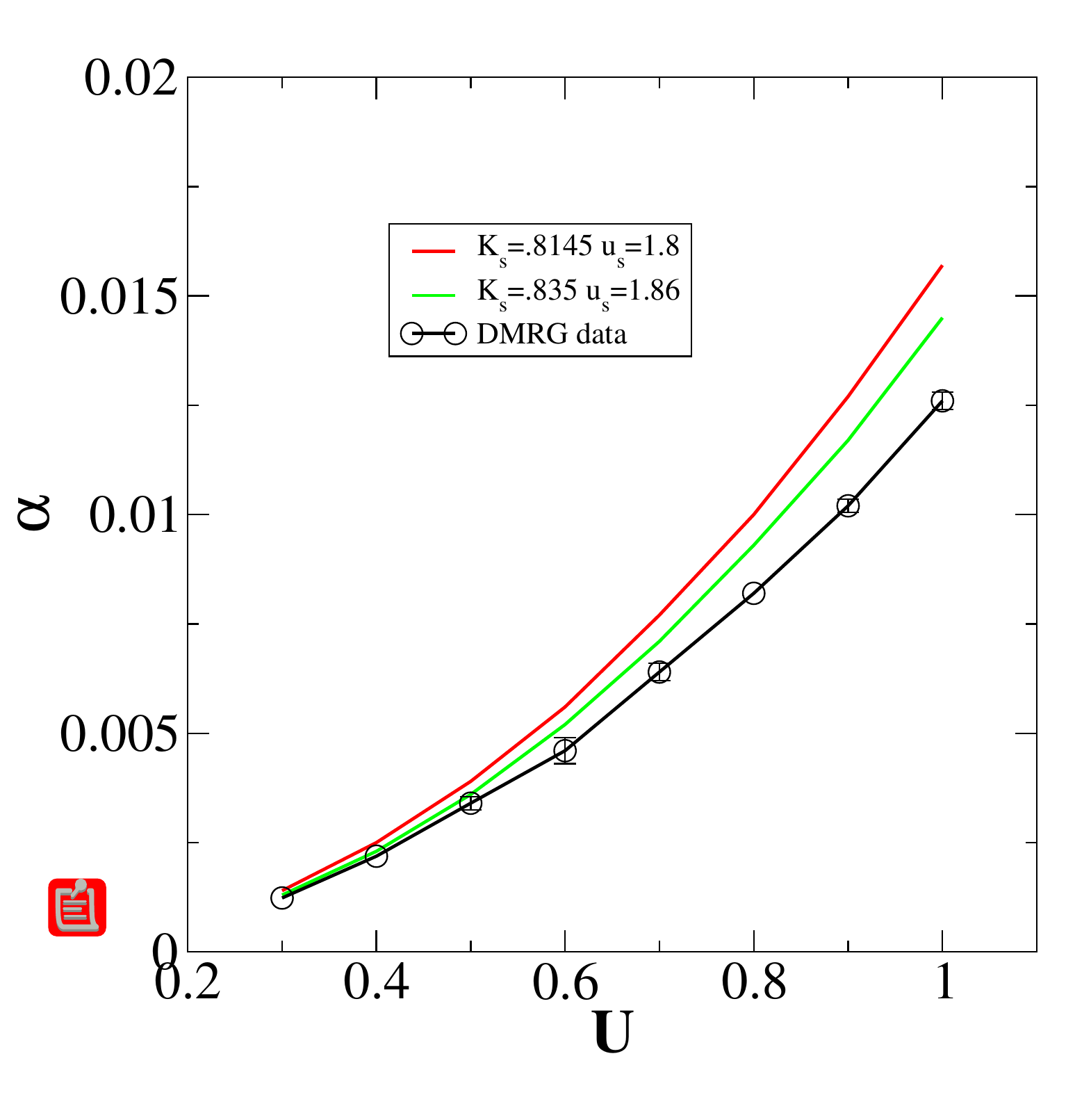}
\end{center}
\caption{\label{fig:expo_p} (color online)   Green's function exponent of the impurity in a ladder of hard core bosons,  as a function of $U$ for  $p =0.1\pi$ (upper panel)$, 0.2\pi$ (lower panel) $t_{\text{imp}}=1$,  transverse hopping  ($t_\perp=1$)  and $\chi=600$. The black curves represent the DMRG data and the green and red curves are fitted curves of the link cluster result with the TLL parameters are mentioned in inset.}
\end{figure}

For small momentum, a momentum dependent exponent can be extracted by the fitting methods described
in Appendix.~\ref{ap:data_extraction}. The results are shown in Fig.~\ref{fig:expo_p}.
The growth of the exponent with momentum can be expected since on general grounds one finds \cite{zvonarev_ferro_cold,kantian_impurity_DMRG} that the exponent varies as
\begin{equation}
 \alpha(p) = \alpha(p=0) + \beta p^2
\end{equation}
The coefficient $\beta$ can be computed for small interactions (see Appendix~\ref{ap:LCE}) and is given by
\begin{equation}
 \beta = \frac{3K_sU^2}{ \pi^2 u_s^2}\frac{ t_{\text{imp}}^2}{u_s^2}
\end{equation}
Comparison between the numerics and the LCE shows a very good agreement between the two for small interactions.

This good agreement with the numerical exponent and the LCE one suggests that, as for the single chain we can use the LCE to estimate the critical momentum $p^*$ which separates the ID regime from the polaronic one.
The crossover depends on the TLL characteristics of the bath, namely the velocity of sound in the ladder and the TLL parameter $K$. Using the values extracted from \cite{crepin_bosonic_ladder_phase_diagram} we get
\begin{equation}
 p^* =0.3\pi
\end{equation}
which is in reasonably good agreement with the observed change of behavior in Fig.~\ref{fig:decay}.
Beyond the LCE we also see that the change of behavior is in good agreement with the criteria of intersection discussed
in Fig.~\ref{fig:Structure_factor} when one uses properly the structure factor the ladder.

\begin{figure}
\begin{center}
  \includegraphics [ scale=0.4]
 {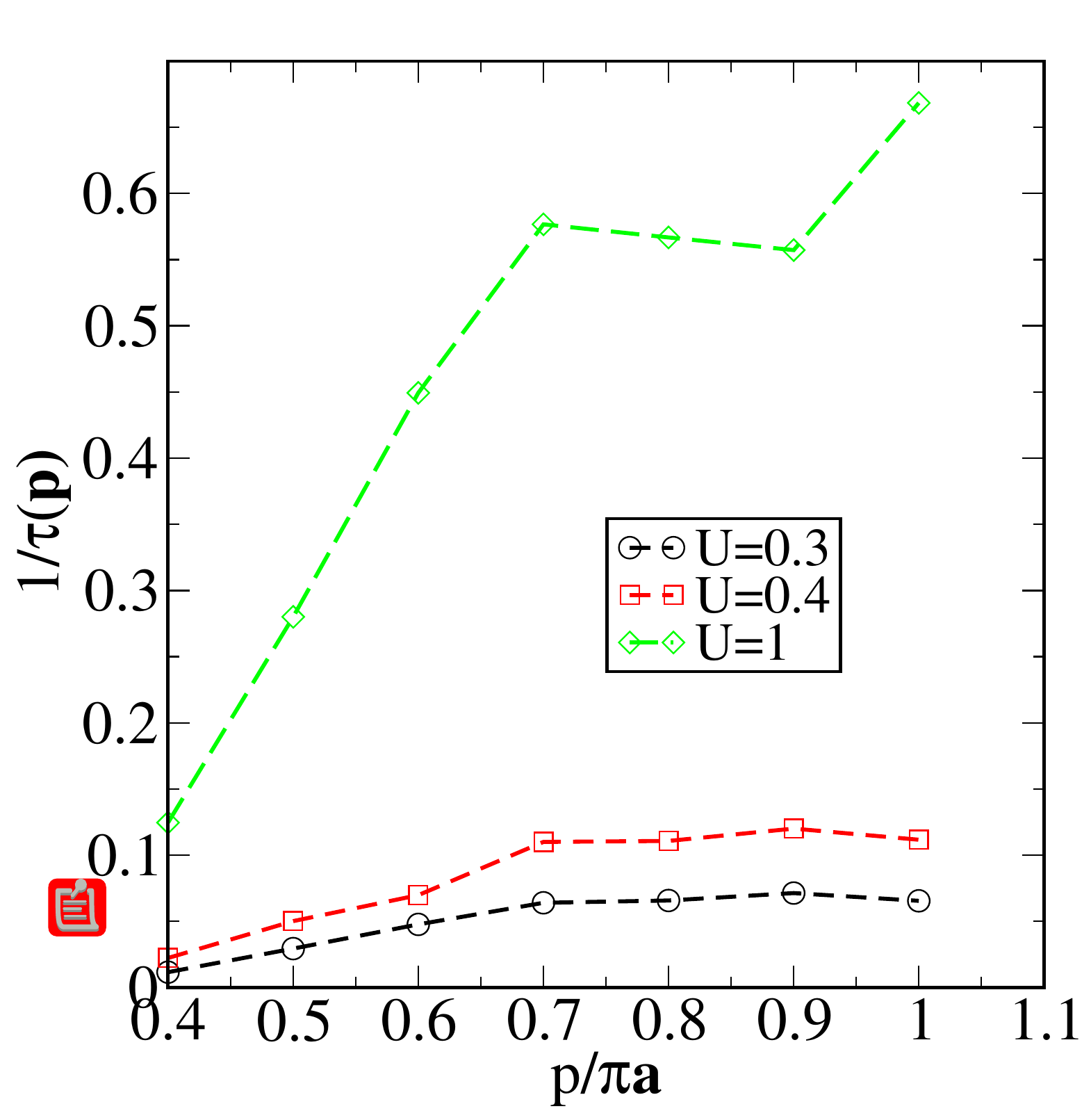}
\end{center}
\caption{\label{fig:life_time_p} (color online) Inverse life time of the impurity as function of the momentum for a bath of hard core bosons at one third filling. The parameters are $t_{\text{imp}}=t_b=t_\perp=1$, $U=0.3, 0.4, 1$, $a=1$ is lattice spacing.}
\end{figure}

\begin{figure}
\begin{center}
  \includegraphics [scale=0.35]
 {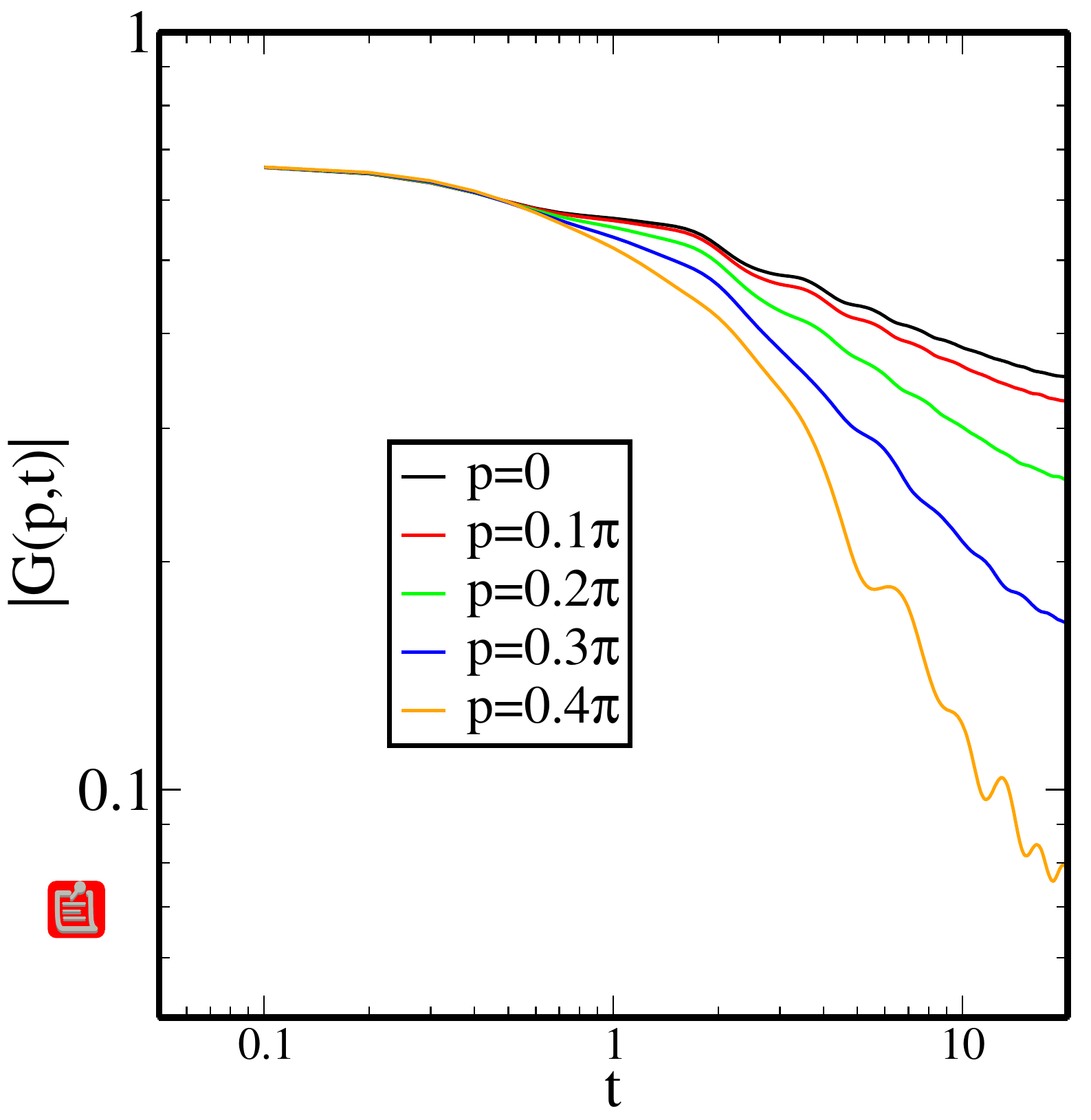}
  \includegraphics [scale=0.35]
 {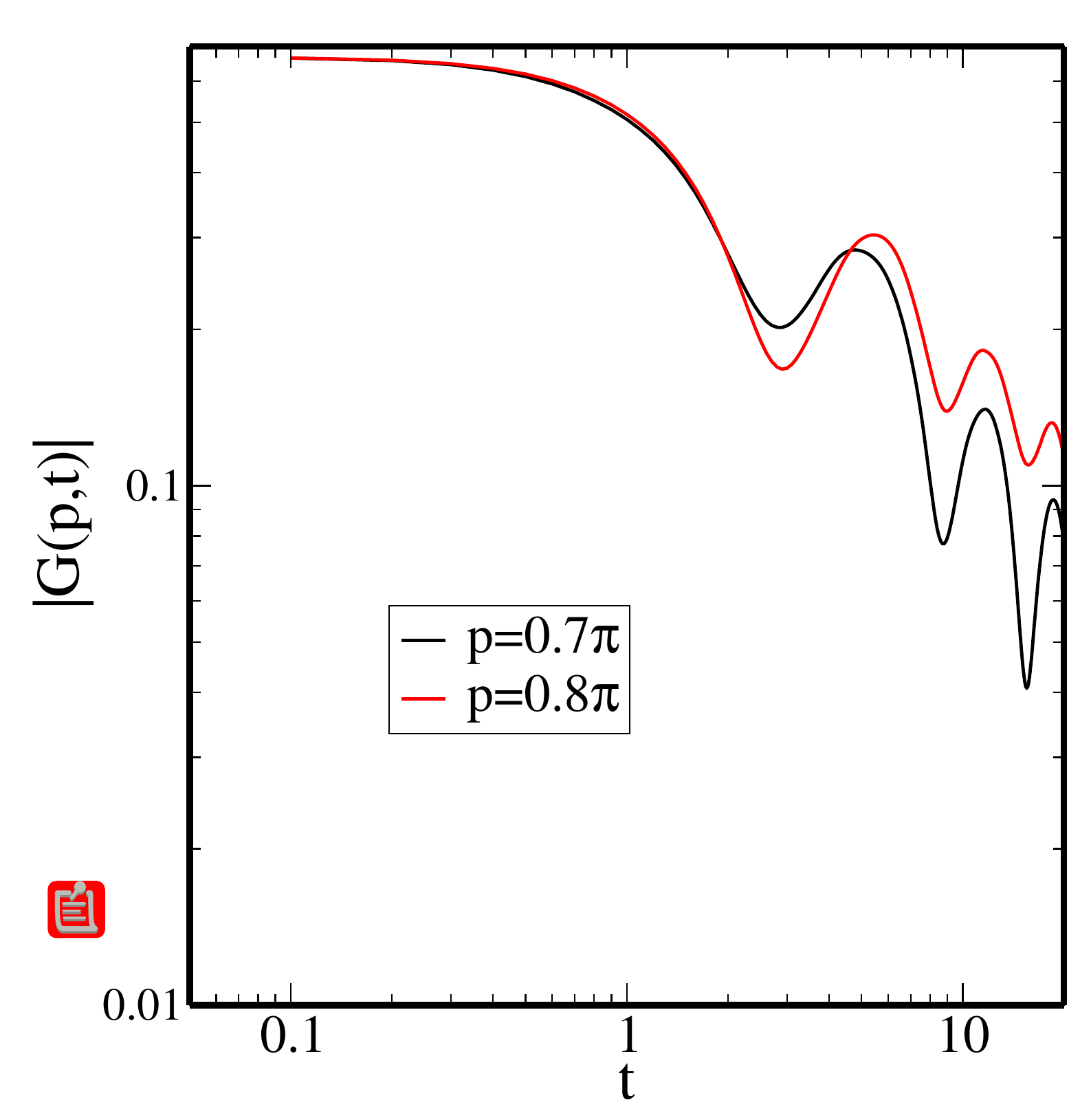}
 \includegraphics [scale=0.35]
{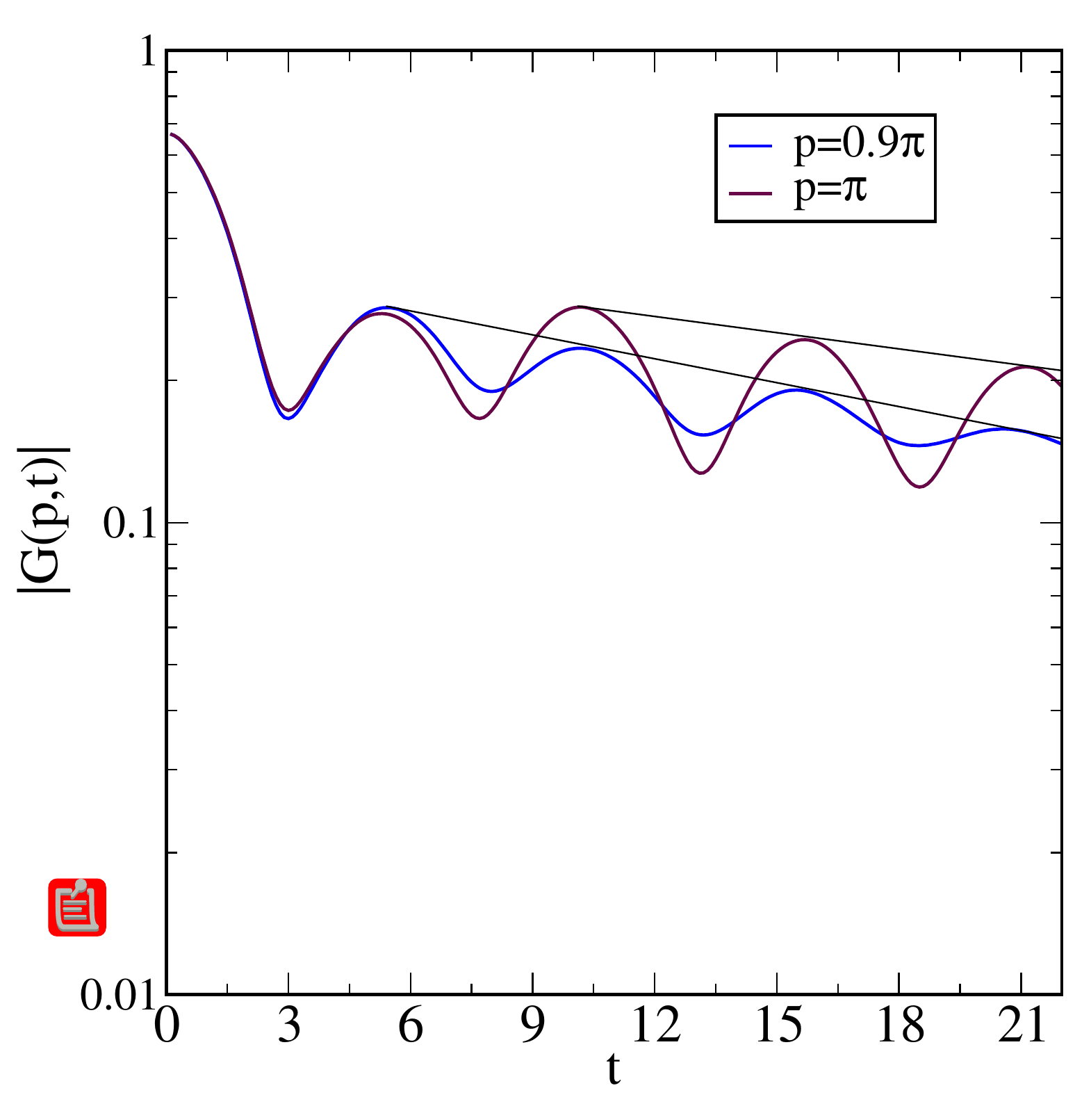}
\end{center}
\caption{\label{fig:2fig6}(color online) Green's function of the impurity as a function of time for $p=0, 0.1\pi, 0.2\pi, 0.3\pi, 0.4\pi, 0.7\pi, 0.8\pi, 0.9\pi, \pi $ for hard core boson on log-log $(p=0 - 0.8\pi)$ scale and semi-log scale $(p=0.9\pi,  \pi)$, $\text{t}_\perp=1.5$, $\text{U=}\infty$, $ t_{\text{imp}}=1$, for bond dimensions $\chi=600$.}
\end{figure}

Beyond $\text{p}=\text{p}^*$ and for small $U$, the Green's function decays exponentially, the impurity behaves like a quasi-particle, and the Green's function of the impurity in term of life time $\tau(p)$ is given by
\begin{equation}\label{eq:life_time}
|G(p,t)|=\exp(-t/\tau(p))
\end{equation}
In Fig.~\ref{fig:life_time_p}, we plot the inverse of life time $1/\tau(p)$, defined in eq. (\ref{eq:life_time})  of the QP as function of $\text{p}$ for different interactions $U=0.3, 0.4, 1$. As can be expected $1/\tau(p)$ increases with increasing interaction.
Let us now turn to the limit of infinite repulsion between the impurity and the bath.
The numerical Green's functions are given in Fig.~\ref{fig:2fig6}.
In a similar way than for the small interaction limit, the numerical data shows a crossover from a powerlaw regime to another type of decay upon increasing the momentum. The transition occurs for after $p$ around $p^* = 0.9\pi$.

\section{Discussion} \label{sec:discussion}

Our results for the ladder system give various indication on how the quantum bath affect the impurity when going from a
strictly one dimensional situation to a higher dimensional one. As usual  going from one dimension to e.g. two
dimensions is not smooth and we also see some direct effects of the ladder structure.

Let us first turn to the weak coupling between the bath and the impurity. The numerical results fully confirm that in such limit
the field theory description (\ref{eq:2eq1}) gives an excellent approximation of the  exponent at zero momentum.
From (\ref{eq:2eq1}) and the numerical results, we see that
the system goes toward a more conventional superfluid, in which only the symmetric mode is massless but the transverse
mode gap leads to a \emph{reduction} of the decay exponent at zero momentum, and roughly replaces the interaction between bath
and impurity by $U/\sqrt2$. All things being equal for the bath in term of TLL parameter $K$ (see below) this indeed leads
to a reduction of the decay exponent. This is well in agreement with the idea that in higher dimension recoil of the impurity
cures the effects of Anderson orthogonality catastrophe. We could easily extrapolate this result to a system of $N$ chains for
which we could expect all the modes except the global symmetric mode to be gapped in which case we would get a reduction of the
interaction of the order of $U/\sqrt{N}$.

We however see that some of the peculiarities of the ladder, which would likely disappear when increasing the number of legs,
manifest themselves in the exponent in addition to the above mentioned effect.
The ladder itself is more affected by the interactions \emph{within} the bath than a single chain.
This is readily seen by the fact that a ladder of hard core bosons at a filling of one boson per rung would be an insulator,
while a single chain remains a superfluid at half filling. As a consequence the TLL parameter $K$ can reach values below one \cite{crepin_bosonic_ladder_phase_diagram}
corresponding to ``enhanced'' repulsion compared to the case of a single chain. This effect contributes also to the decay
and to a further reduction of the exponent compared to the effect on $U$ itself.

On the other hand increasing $t_\perp$ leads to an opposite trend, particularly visible in the limit of large $t_\perp$ (see Fig \ref{fig:2fig5} ), in which a marked \emph{increase} of the exponent can be seen.
Naively one could consider that increasing $t_\perp$ brings back the system to a single mode for the rung and thus pushes the system back to a more one-dimensional behavior. 
However the impurity couples still to both modes making the calculation of the exponent delicate.
How such effects would be modified upon increasing the number of legs is of course an interesting and open question.

As can be seen from Fig.~\ref{fig:interaction}  the general expression that was introduced in \cite{kantian_impurity_DMRG} provides a good
description of the decay exponent at zero momentum for the ladder as well. Note that in the present case we have a full numerical evaluation of the infinite repulsion case between impurity and the bath, making it an essentially parameter free formula and allowing
to check that the large impurity-bath repulsion limit is indeed correctly reproduced by this formula.
This confirms that for a system in which the impurity couples to a single massless mode the generalization of the free fermion solution (\ref{eq:empform}) captures the essential physics, and we would expect it to hold to the case of $N$ chains
as well. As for the single chain, the finite momentum case remains more difficult to interpret. We recover for the ladder the same
two regimes that were observed in the single chain, namely a regime dominated by infrared divergences (ID) for which we have
the powerlaw decay of the correlations, and a different regime, which seems, within the accuracy of the numerical solution
also correspond to a faster decay of the correlations, and thus to a more conventional polaronic regime.
As for the single chain the separation between these two regimes seems to be reasonably well given by the possibility to
excite real particle-hole excitations in the bath, as shown by Fig.~\ref{fig:Structure_factor}.

\section{Conclusion and perspectives} \label{sec:conclusion}

We have studied the effect of a quantum bath made of a bosonic ladder on an impurity confined to move in one dimension.
We have computed, both analytically and by using time dependent density matrix renormalization group,
the Green's function of the impurity as a function of the interaction between the impurity and the bath
and the interactions in the bath itself.

We find that, as for a single chain bath, the presence of the bath affects drastically the mobility of the impurity, in
a way quite different than a simple renormalization of the mass of the impurity via a polaronic effect.
One measure of such effects is given by the Green's function of the impurity corresponding to the creation of an impurity
with a given momentum $p$ at time zero and its destruction at time $t$. We compute this Green's function both analytically
using a bosonization representation of the bath, a linked cluster expansion and numerically with a time
dependent density matrix renormalization group technique.

We find that for small momentum the Green's function decays as a power law. For weak interaction between the impurity
and the bath, the exponent is smaller than for a single chain, due to the gap appearing in the antisymmetric mode of the ladder.
This trend would increase with the number of legs of the ladder, and reflect the trend to the more two dimensional behavior.
On the other hand some aspects of the ladder are manifest in the fact that for large tunnelling in the ladder one finds an increase
of the exponent.

We studied also the dependence of the Green's function on the momentum of the impurity and found, in a similar way that for the
single chain that there is another regime than the power-law decay that appears, in which one has an exponential decay.
The transition between these two regimes is well connected with the possibility to excite real rather than virtual
particle hole excitations in the bath.

Our analysis is a first step towards studying the evolution from the 1D behavior, essentially dominated by the
Anderson orthogonality catastrophe provoked by the motion of the impurity, and which in 1D is not cured by the motion itself,
and the higher dimensional behavior, in which more conventional polaronic behavior is expected. Several additional directions
could prove interesting extensions. If the interaction in the bath becomes sufficiently strong and sufficiently long
range, backscattering on the bath becomes as relevant as the forward scattering considered in the present study. It would
be interesting to know the effect on the impurity in ladders. Another extension could be to let the impurity
also delocalize on the various chains so that the motion of the impurity itself tends to a two dimensional one.

Cold atomic systems could provide a good realization of the systems described in the present paper. On one hand bosonic
ladders have been realized \cite{atala_ladders_meissner} and more generally systems made of many bosonic tubes are routinely
realizable either in bulk \cite{stoferle_tonks_optical} or in boson microscopes. Atom chips systems provide
also an excellent realization of a ladder system \cite{hofferberth_interferences_atomchip_LL}. Systems with bath and impurities
have already been realized for single chains \cite{Michael_Kohl_Spin_impurity_in_bose_gas,Minardi_Dynamics_of_impurities_in_one_dimension,%
Fukuhara_spin_impurity_in_one_dimension,meinert_bloch_oscillations_TLL} either by using two species (such as K and Rb) or internal
degrees of freedom. Combination of these two aspects should be reachable in a very near future.

\acknowledgments
Calculations were performed using the Matrix Product Toolkit~\cite{mptoolkit}.
We thank N. Laflorencie and G. Roux for providing us with the precise numerical value for the TLL parameters of the ladder of publication
\cite{crepin_bosonic_ladder_phase_diagram}. This work was supported in part by the Swiss NSF under division  II.

\appendix

\section{Linked Cluster Expansion}\label{ap:LCE}

We have shown in the main text that for small interaction between  impurity and  ladder compared to  the gap in  the antisymmetric sector, the impurity effectively couples to the forward scattering part of the symmetric sector.
We give in this appendix the LCE calculation of the Green's function of the impurity for a weak interaction $U$. We consider the symmetric part of (\ref{eq:2eq7}) as the bath Hamiltonian, and represent it in term of the usual bosonic operators~\cite{Giamarchi_Bosonization}.

In second quantized notation (\ref{eq:effhamiltonian}) is described by
\begin{eqnarray}
H&=&H_s+ H_\text{imp}+H_\text{coup}\nonumber \\
H_s&=&\sum_{q}u_s|q| b^\dagger_{s q}b_{s q}\nonumber \\
H_\text{coup}&=&\sum_{{q}, {k}}V(q) d^\dagger_{{ k}+{q}}d_{{ k}}(b_{{s q}}+b^\dagger_{{s -q}})\nonumber \\
H_\text{imp}&=&\sum_{q}\epsilon(q) d^\dagger_{q}d_{ q}\nonumber \\
\label{eq:2eq1A}
\end{eqnarray}
Where $b^\dagger$ and $d^\dagger$ are the creation operators for the bath and the impurity respectively, $H_\text{imp}$ is the tight binding Hamiltonian of the impurity , and $H_\text{coup}$ is interaction between impurity and bath.
\begin{eqnarray}
V(q)&=&\frac{U}{\sqrt 2} \sqrt{\frac{K_s|q|}{2\pi L}}\exp\Big(-\frac{|q|}{2q_c}\Big)
\label{eq:2eq2A}
\end{eqnarray}
where $q_c$ is an ultraviolet cutoff of the order of the inverse lattice spacing.

The Green's function of the impurity is defined as
\begin{eqnarray}
G(p,t)&=&-i\langle d_{ p}(t)d^\dagger_{ p}(0)\rangle
\label{eq:2eq3A}
\end{eqnarray}
By using LCE, (\ref{eq:2eq3A}) can be written as
\begin{eqnarray}
G(p,t)&=&-ie^{-i\epsilon_{p}t}e^{F_2(p,t)}
\label{eq:2eq4A}
\end{eqnarray}
Where $F_2(p,t)$ is defined as
\begin{eqnarray}
F_2(p,t)&=&e^{i\epsilon_{p}t}W_2(p,t)
\label{eq:2eq5A}
\end{eqnarray}
$W_2(p,t)$ is given by
\begin{eqnarray}
W_2(p,t)&=&-\frac{1}{2}\int_{0}^{t} dt_1\int_{0}^{t} dt_2\nonumber\\&& \langle T_\tau d_{p}(t) H_{coup}(t_1)H_{coup}(t_2) d^\dagger_{p}(0) \rangle\nonumber\\
\label{eq:2eq6A}
\end{eqnarray}
By employing Wick's theorem $W_2(p,t)$ is given by
\begin{eqnarray}
W_2(p,t)&=-&\sum_{q}V(q)^2\int_{0}^{t} dt_1\int_{0}^{t} dt_2 Y(t_1) e^{-i\epsilon(p)t_1}\nonumber\\&&Y(t_2-t_1) e^{-i\epsilon(p+q)(t_2-t_1)} Y(t-t_2)\nonumber\\&&e^{-i\epsilon(p)(t-t_2)}Y(t_2-t_1)e^{-i(u_s|q|(t_2-t_1))}\nonumber\\&&
\label{eq:2eq7A}
\end{eqnarray}

Where $Y(t)$ is a step function, which is zero for $t<0$ and one for $t>0$. $Y(t)$ changes the limit of integration of $t_2$ and $t_1$, and $F_2(p,t)$ is modified as
\begin{eqnarray}
F_2(p,t)&=-&\sum_{q}V(q)^2\int_{0}^{t} dt_2\int_{0}^{t_2} dt_1  e^{-i\epsilon(p)t_1}\nonumber\\&& e^{-i\epsilon(p+q)(t_2-t_1)} \nonumber\\&&e^{-i\epsilon(p)(t-t_2)}e^{-i(u_s|q|(t_2-t_1))}\nonumber\\&&
\label{eq:2eq9A}
\end{eqnarray}
\begin{eqnarray}
F_2(p,t)&=&-\sum_{q}\int du V(q)^2\int_{0}^{t} dt_2\int_{0}^{t_2} dt_1 \nonumber\\&& e^{-it_1u} e^{it_2 u}\nonumber\\&&\delta(u-(\epsilon(p)-\epsilon(p+q)-u_s|q|))\nonumber\\&&
\label{eq:2eq9B}
\end{eqnarray}
After performing an integration over $t_1$ and $t_2$, $F_2(p,t)$ is given by
\begin{eqnarray}
F_2(p,t)&=&-\sum_{q}\int du V(q)^2 \nonumber\\
&&\frac{1+iut-e^{itu}}{u^2}\nonumber\\&&\delta(u-(\epsilon(p)-\epsilon(p+q)-u_s|q|))
\label{eq:2eq10A}
\end{eqnarray}
which can be rewritten as
\begin{equation}\label{eq:2eq11A}
F_2(p,t)= -\int du  \frac{1+iut-e^{itu}}{u^2} R(u)
\end{equation}
For small momentum $\epsilon(p)\simeq t_{\text{imp}} p^2$,
\begin{equation}\label{eq:2eq12A}
R(u)= \sum_{q}V(q)^2\delta(u-(\epsilon(p)-\epsilon(p+q)-u_s|q|))
\end{equation}

We evaluate $R(u)$ as
\begin{equation}
R(u) = \frac{1}{2\pi}\int dqV(q)^2\delta(u-(\epsilon(p)-\epsilon(p+q)-u_s|q|))
\label{eq:2eq13A}
\end{equation}
$q$'s roots inside the delta function are given by
\begin{equation}\label{eq:2eq14A}
q{^>_\pm} = -\Big(p+\frac{u_s}{2t_{\text{imp}}}\Big) \pm \sqrt{\Big(p+\frac{u_s}{2J}\Big)^2-\frac{u}{t_{\text{imp}}}}
\end{equation}
and
\begin{equation}\label{eq:2eq15A}
q{^<_\pm} =-\Big(p-\frac{u_s}{2t_{\text{imp}}}\Big) \pm \sqrt{\Big(p-\frac{u_s}{2t_{\text{imp}}}\Big)^2-\frac{u}{t_{\text{imp}}}}
\end{equation}
$q^>$ corresponds to $q>0$ and $q^<$ corresponds to $q<0$.
For $(p-\frac{u_s}{2t_{\text{imp}}})<0 \wedge u>0$, $R(u)=0$ while
for  $(p-\frac{u_s}{2t_{\text{imp}}})>0 \wedge u>0$ one has
\begin{multline}\label{eq:2eq16A}
R(u) = \frac{1}{2t_{\text{imp}}}\Big[\Big(\frac{p-u_s/(2t_{\text{imp}})}{\sqrt{{(p-u_s/(2t_{\text{imp}}))}^2-u/t_{\text{imp}}}}-1\Big) \\
  e^{-|q^<_{+}|/(2q_c)}+ \Big(\frac{p-u_s/(2t_{\text{imp}})}{\sqrt{{(p-u_s/(2t_{\text{imp}}))}^2-u/t_{\text{imp}}}}+1\Big) \\
  e^{-|q^<_{-}|/(2q_c)}\Big]
\end{multline}
For $u<0$.
\begin{multline}\label{eq:2eq17A}
R(u) = \frac{1}{2t_{\text{imp}}}\Big[\Big(1-\frac{p+u_s/(2t_{\text{imp}})}{\sqrt{{(p+u_s/(2t_{\text{imp}}))}^2-u/t_{\text{imp}}}}\Big) \\
 e^{-|q^>_{+}|/(2q_c)}+\nonumber\\ \Big(\frac{p-u_s/(2t_{\text{imp}})}{\sqrt{{(p-u_s/(2t_{\text{imp}}))}^2-u/J}}+1\Big)e^{-|q^<_{-}|/(2q_c)}\Big]
\end{multline}
If $(p-\frac{u_s}{2t_{\text{imp}}})<0,  \wedge  u<0$
\begin{eqnarray}
R(u)&\simeq&u
\label{eq:2eq18A}
\end{eqnarray}
\begin{eqnarray}
Re[F_2(p,t)]&\simeq&-\log(t)
\label{eq:2eq19A}
\end{eqnarray}
For small $|(p-\frac{u_s}{2 t_{\text{imp}}})|$
\begin{equation}\label{eq:2eq20A}
Re[F_2(p,t)] \simeq -\frac{K_sU^2}{4 \pi^2 u_s^2}(1+\frac{12 t_{\text{imp}}^2p^2}{u_s^2}) \log(t)
\end{equation}
leading to the Green's function decay
\begin{eqnarray}
|G(p,t)|&=&e^{-\frac{K_sU^2}{4 \pi^2 u_s^2}(1+\frac{12 t_{\text{imp}}^2p^2}{u_s^2})\log(t)}
\label{eq:2eq21A}
\end{eqnarray}

\section{DMRG procedure} \label{ap:dmrg}

\begin{figure}
\begin{center}
  \includegraphics [scale=0.5]{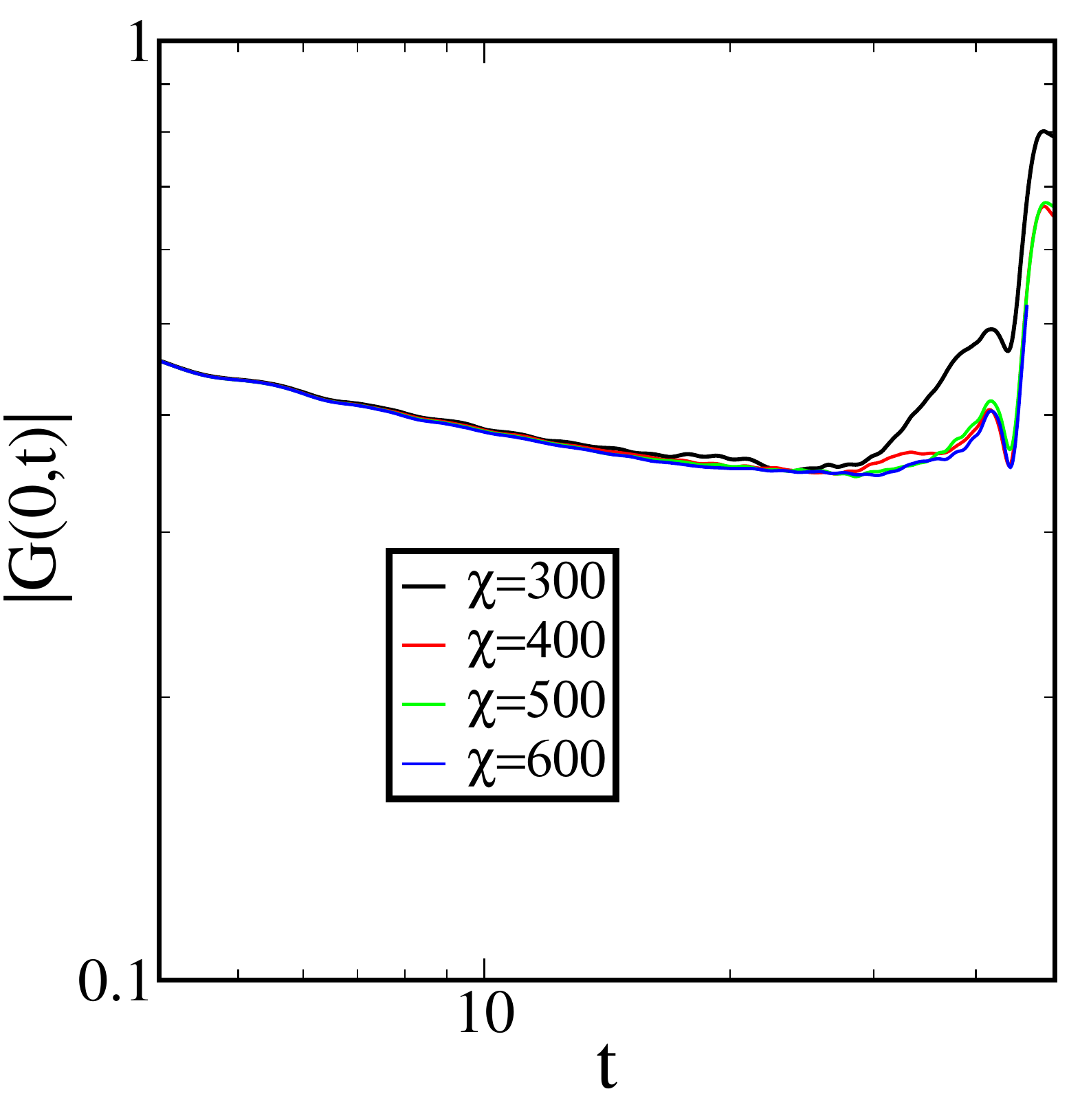}
\end{center}
\caption{\label{fig:2fig7} (color online) Green's function as a function of time at $\text{p=0}$ for hard core boson, $\text{t}_\perp=1.5$, $\text{U=}\infty$, $\text{t}_b=1$, $t_{\text{imp}}=1$
for bond dimensions $\chi=300,400,500,600$. With increasing $\chi$ the upturn in Green's function pushed down.}
\end{figure}

In Fig.~\ref{fig:2fig7}, we show the Green's function of the impurity for $\chi=300,400,500,600$ at zero momentum. We find that the Green's function upturns after a certain time. This is because entanglement entropy linearly scale with time, so one needs exponentially large $\chi$ to get a good result. We find that with increasing $\chi$, the upturn which is a numerical artifact that signals the growth of entanglement entropy with time evolution, is pushed down.

\section{Extracting exponent and error bar from numerical data}\label{ap:data_extraction}

The absolute value of Green's function of the impurity decays monotonically as a function of time but has also oscillations. To  extract the exponents, we first select all the data (up to a maximum time) where Green's function is maximum. We compute the slope between two neighboring data and  take the average of all the slopes to obtain the exponent.

If on a log-log scale, $\log|G(0,t_l)|'s$ are maxima of $\log|G(0,t)|$, at time $t_l$, where $l=1,2,..n$,
then the slope between time $t_{l}$ and $t_{l+1}$ is given by $\beta_l=\frac{\log(|G(0,t_l))-\log(|G(0,t_{l+1}))}{\log(t_{l+1})-\log(t_{l})}$, and the exponent $\alpha=\frac{\beta_1+\beta_2+.....\beta_{n-1}}{n-1}$.  The error bar is given by $\frac{|\alpha-\beta_1|+|\alpha-\beta_2|+.....+|\alpha-\beta_{n-1}|}{n-1}$.

\begin{figure}
\begin{center}
  \includegraphics [ scale=0.25]
 {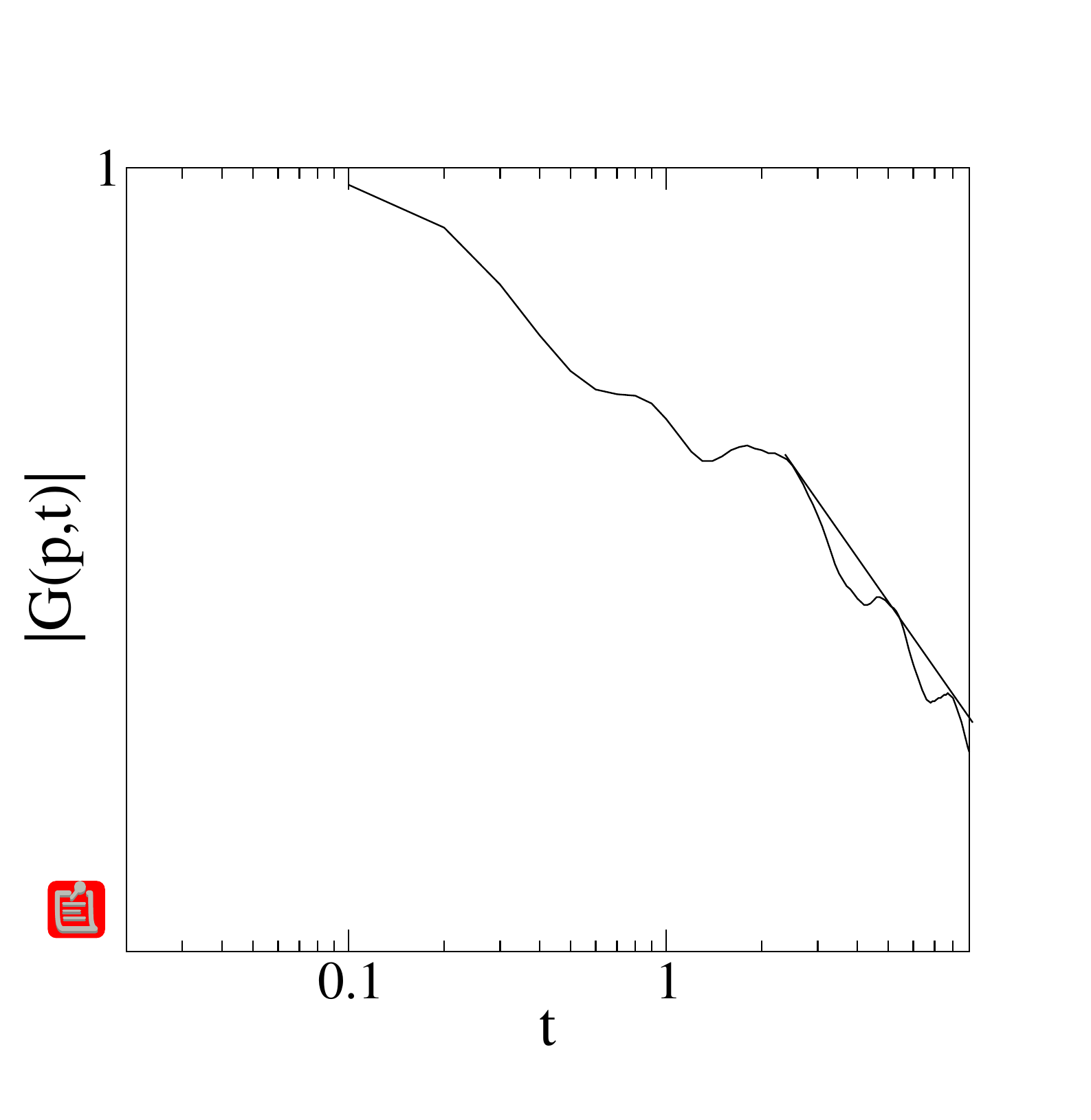}
  \includegraphics [ scale=0.25]{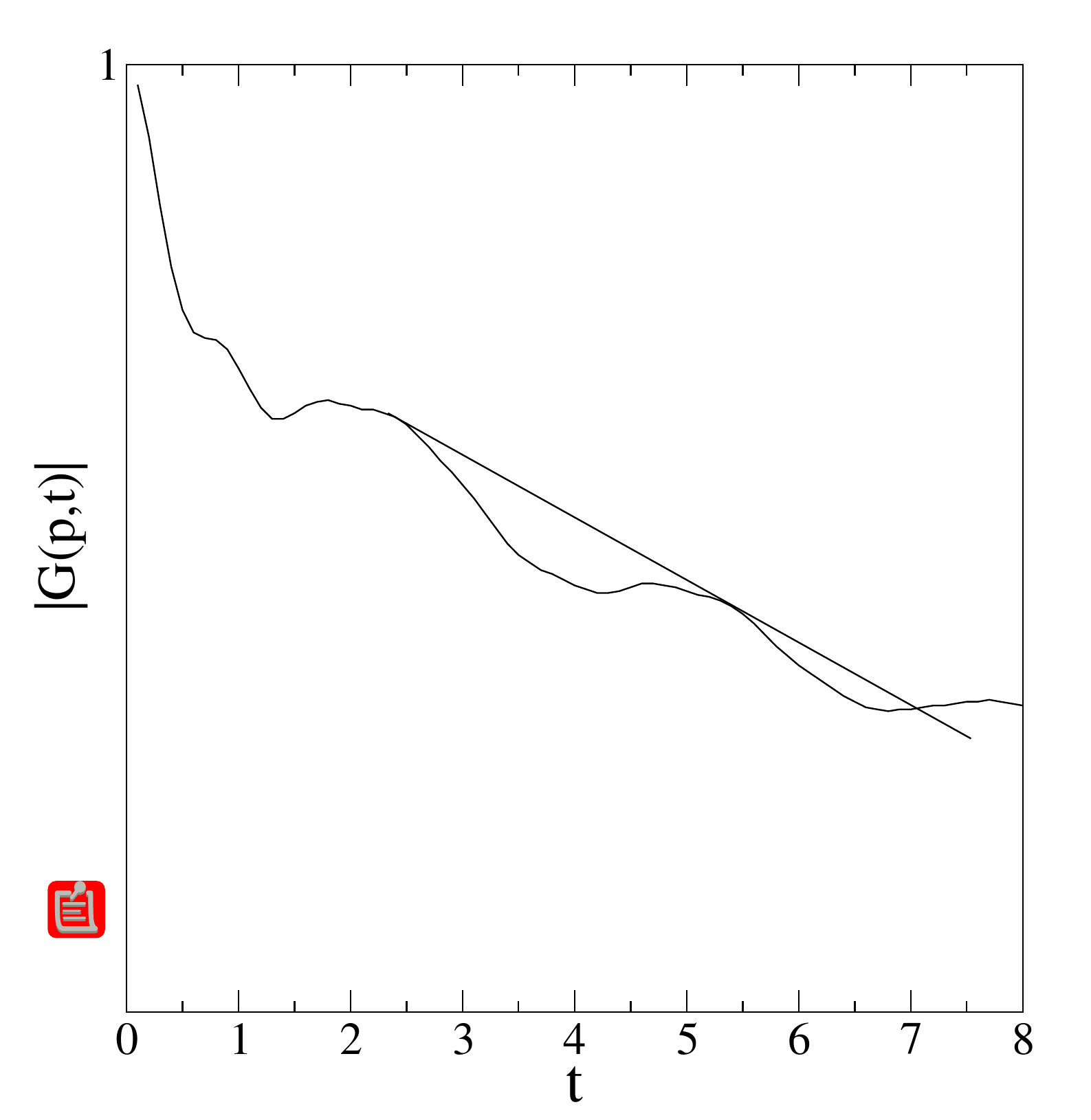}
   \includegraphics [ scale=0.25]
 {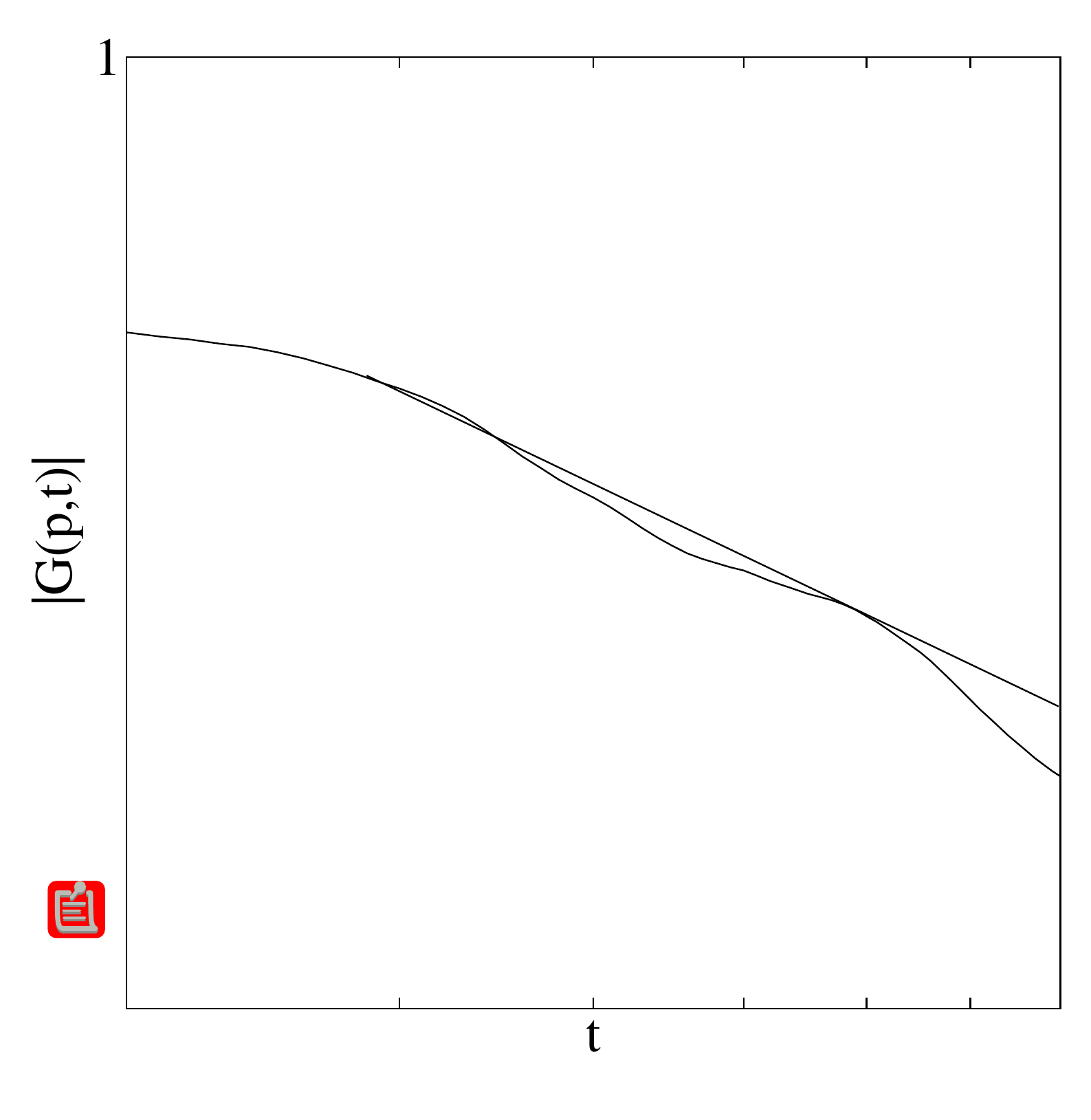}
  \includegraphics [ scale=0.25]{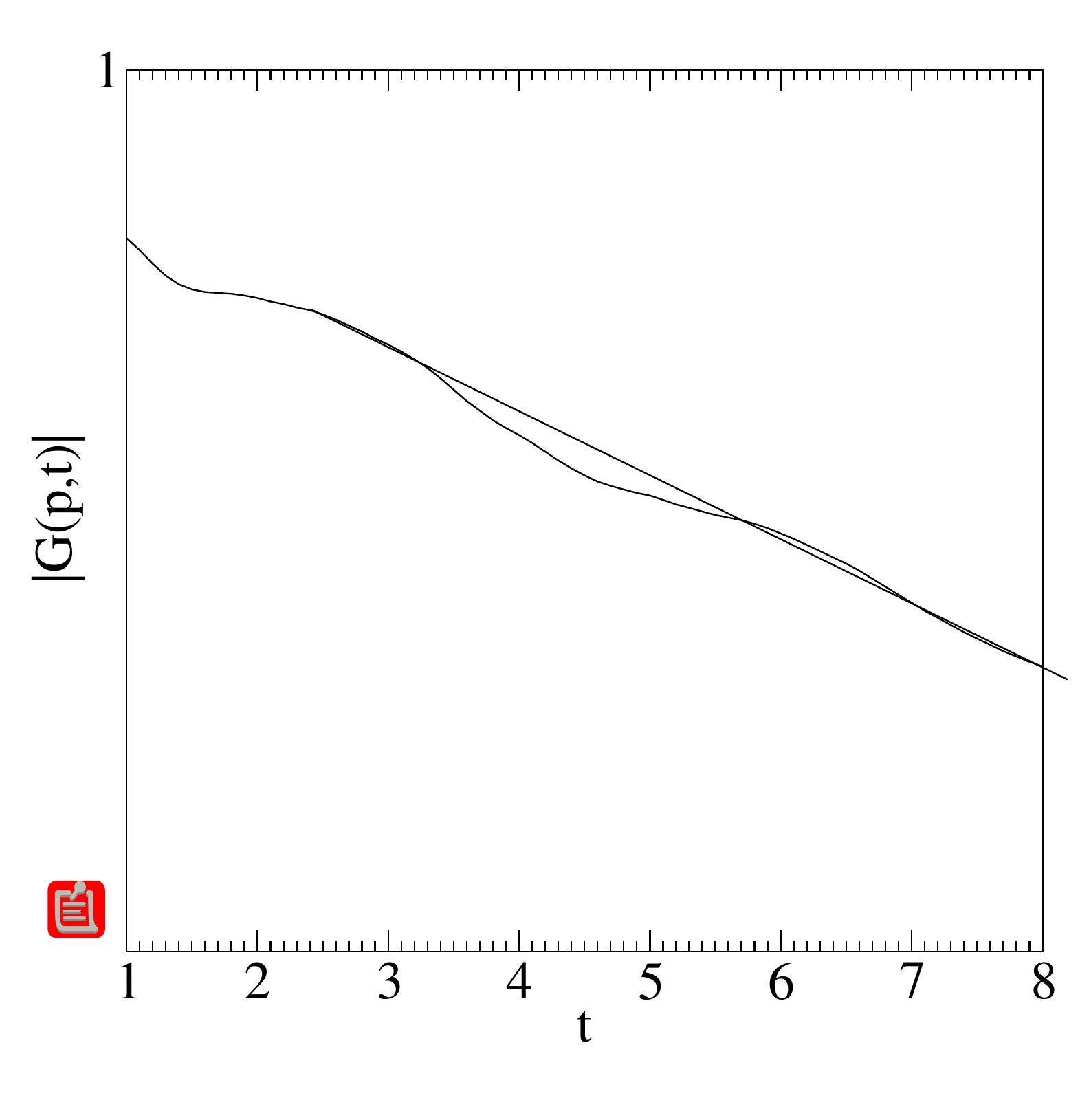}
\end{center}
\caption{\label{fig:2fig8} (color online)  Green's function of the impurity as function of time for hard core bosons at $t_\text{imp}=t_\text{b}=1$,  $\text{t}_\perp=1$, $\text{U}=1$. The upper panel represents the Green's function at $\text{p}=0.2\pi$, and the lower panel the Green's function at $\text{p}=0.3\pi$  on log-log scale (left) and semi-log scale (right). At $\text{p}=0.2\pi$, the Green's function is linear on a log-log scale but deviates on a semi-log scale from linear behavior. On the other hand at
$\text{p}=0.3\pi$, the Green's function is linear on a semi-log scale but deviates on a log-log scale from linear behavior. This allows
us to fix $p^* \sim 0.3 \pi$ as the critical momentum at which the decay for the impurity goes from powerlaw to exponential.  }
\end{figure}

To distinguish between a power-law decay and an exponential one, and determine the critical momentum $p^*$ at which such a
change occurs, we show in Fig.~\ref{fig:2fig8} the Green's function of impurity at $\text{U}=1$, $\text{p}=0.2\pi$ on both
log-log scale and semi-log scale.

The comparison both on a log-log scale and a semi-log one of the these two behaviors allow us to determine $p^* \sim 0.3 \pi$ at which the change of behavior from powerlaw to exponential occurs. At $\text{p}=0.2\pi$, the Green's function decays but has oscillations. To extract the exponent we select all the points where the Green's function is maximum and use the above procedure to compute the error bars and exponent.
\bibliography{totphys-cleaned,Impurity_in_Ladder,impurity_TG}

\end{document}